\pdfoutput=1
\documentclass[journal]{IEEEtran}
\usepackage{graphicx}
\usepackage{array}
\usepackage{mdwmath}
\usepackage{mdwtab}
\usepackage{eqparbox}
\usepackage{algorithmic}
\usepackage[cmex10]{amsmath}
\usepackage{setspace}
\usepackage{algorithm,caption}
\usepackage{empheq}
\usepackage{cases}
\usepackage{soul}
\usepackage{color}
\usepackage{cite}
\usepackage{amssymb}
\usepackage{epstopdf}
\usepackage{amsmath}
\usepackage{ntheorem}
\usepackage{dsfont}
\usepackage{bbm}
\usepackage{subfig}
\usepackage{stfloats}
\usepackage{mathrsfs}
\allowdisplaybreaks
\ifCLASSINFOpdf
\else
\fi
\begin{document}
\renewcommand{\theenumi}{\roman{enumi}}
%
\title{Differential Modulation for Short Packet Transmission in URLLC}
\author{Canjian~Zheng, Fu-Chun~Zheng,~\IEEEmembership{Senior Member,~IEEE}, Jingjing~Luo,~\IEEEmembership{Member,~IEEE}, Pengcheng~Zhu,~\IEEEmembership{Member,~IEEE}, Xiaohu~You,~\IEEEmembership{Fellow,~IEEE} and Daquan~Feng,~\IEEEmembership{Member,~IEEE}}
%
%
\maketitle


\begin{abstract}
One key feature of ultra-reliable low-latency communications (URLLC) in 5G is to support short packet transmission (SPT). However, the pilot overhead in SPT for channel estimation is relatively high, especially in high Doppler environments. In this paper, we advocate the adoption of differential modulation to support ultra-low latency services, which can ease the channel estimation burden and reduce the power and bandwidth overhead incurred in traditional coherent modulation schemes. Specifically, we consider a multi-connectivity (MC) scheme employing differential modulation to enable URLLC services. The popular selection combining and maximal ratio combining schemes are respectively applied to explore the diversity gain in the MC scheme. A first-order autoregressive model is further utilized to characterize the time-varying nature of the channel.
Theoretically, the maximum achievable rate and minimum achievable block error rate under ergodic fading channels with PSK inputs and perfect CSI are first derived by using the non-asymptotic information-theoretic bounds.
The performance of SPT with differential modulation and MC schemes is then analysed by characterizing the effect of differential modulation and time-varying channels as a reduction in the effective SNR.
Simulation results show that differential modulation does offer a significant advantage over the pilot-assisted coherent scheme for SPT, especially in high Doppler environments.
\end{abstract}

%
%


\IEEEpeerreviewmaketitle
\section{Introduction}
The 5th generation (5G) wireless communications networks support three key service categories: enhanced mobile broadband, massive machine-type communications, and ultra-reliable and low-latency communications (URLLC).
Among these three categories, URLLC covers a variety of mission-critical applications such as factory automation, tactile internet, smart transportation and remote control, and thus has received tremendous attention from both the industrial and the academic communities \cite{she2021PIEEE}.
According to the 3rd generation partnership project (3GPP), a general URLLC service under 5G networks should achieve an end-to-end latency of less than 10 ms and block error rate (BLER) of no more than $10^{-5}$ while under B5G and 6G it should be as low as 1 ms and $10^{-7}$, respectively.

To meet such stringent URLLC performance requirements, it is essential to employ a short packet size, such as that of tens to hundreds of bytes, to reduce latency in frame alignment, transmission and detection.
However, when the packet size is short, the pilot or training symbol overheads for acquiring accurate channel state information may occupy a relatively larger portion in a packet and thus becomes a heavier overhead, which is different from the case of traditional long packet transmission (much larger than 1000 bytes).
In long packet transmission, the amount of information data is far beyond the amount of pilot signal so that the overhead for channel estimation can be commonly ignored or compensated.
Additionally, using pilots in short packet transmission (SPT) may potentially incur a high computational complexity and lead to significant processing delay, especially in multi-node or multi-antenna systems.
As a result, if the conventional pilot based channel estimation methods are used in SPT, the system spectral efficiency and latency performance may degrade significantly \cite{Durisi2016Toward}.

In order to avoid the overhead and latency caused by channel estimation, an alternative scheme is to apply differential modulation in SPT.
With differential modulation, the transmitter is able to encode the information as the phase difference between two consecutive information symbols and the receiver can extract the data by directly comparing the phases of the two successively received symbols\cite{Wang2012Dispensing}.
Therefore, the processes of pilot transmission and channel estimation (both consuming time resource) are no longer required in differential modulation based SPT and in the mean time all the related issues are avoided too.
However, relying solely on differential modulation based SPT may not be sufficient to enable URLLC due to the stringent reliability requirements (i.e., the 3 dB loss by differential modulation may degrade the reliability performance).
Therefore, it is necessary to integrate differential modulation with other effective diversity schemes to compensate or improve the reliability performance of SPT under URLLC (equivalently, by exchanging "space diversity" for "time").



Recently, multi-connectivity (MC) has been treated as one of the most promising diversity schemes for supporting reliable transmission in 5G and B5G systems, where the identical information data is transmitted in a single time slot via multiple communication links \cite{Pupiales2021Multi}.
An example of MC schemes for uplink URLLC is illustrated in Fig. 1.
It has been shown in the existing literature, such as \cite{Suer2020Multi, Wolf2019How, Popovski2019Wireless}, that MC does yield considerable performance improvement in terms of outage probability, coverage and throughput compared to the case of single-connectivity, and thus is suitable for low latency communications.
To our best knowledge, however,
the benefits of jointly employing differential modulation and MC schemes for SPT have rarely been reported in the literature so far, which is the main motivation of this paper.
\begin{figure}[!htp]
\centering
\includegraphics [width=0.45\textwidth]{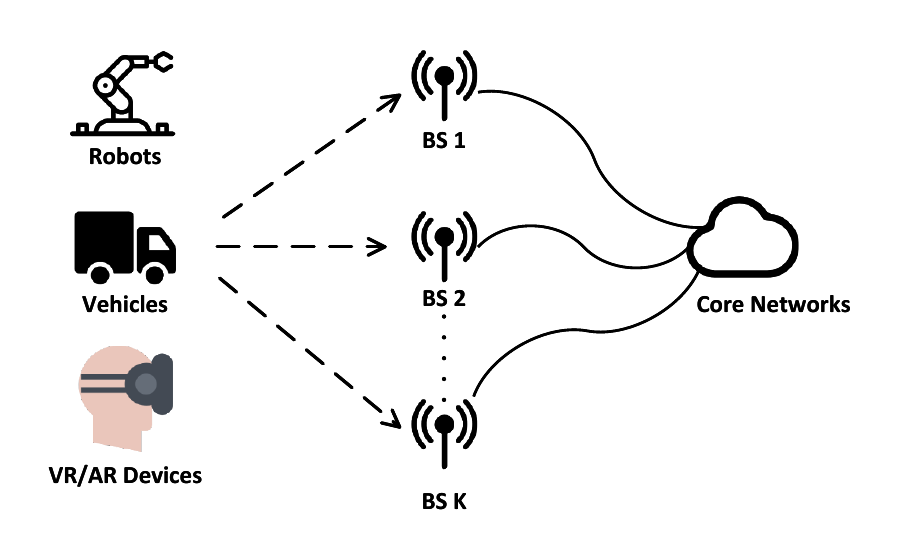}
\caption{A multi-connectivity (MC) scheme where each URLLC device is simultaneously connected to multiple BSs.}%
\label{MC_model}
\end{figure}
\subsection{Related Works}
1) Pilot-assisted SPT:
The performance and pilot overhead optimization of SPT have been extensively investigated in the literature.
Specifically, in \cite{Johan2019Short} and \cite{Ferrante2018Pilot},
the performance of pilot-assisted SPT over the Rician block-fading channels and Rayleigh fading channels was studied by using the non-asymptotic information-theoretic bounds developed in \cite{Polyanskiy2010Channel}.
In \cite{Johan2021URLLC} and \cite{Lancho2023Cell}, a framework that included imperfect channel state information (CSI), pilot contamination, and spatially correlated channels was developed to analyze the packet error probability of pilot-assisted SPT in massive multiple-input multiple-output (MIMO) and cell-free massive MIMO systems.
In \cite{Kislal2023Efficient}, a numerically efficient method was proposed to accurately evaluate the error probability of massive MIMO with pilot-assisted SPT over ergodic Rayleigh fading channels,
while in \cite{Hu2018Finite}, the BLER performance of SPT under cooperative systems was numerically evaluated by considering channel estimation overhead.
Moreover, the pilot overhead optimization of short packets was studied in \cite{Mousaei2017Optimizing} for a point-to-point communications system.
Furthermore, joint optimization of pilot length and block length in SPT was investigated in \cite{Cao2022Independent} for a single-device system and in \cite{Ren2020Joint} for massive MIMO systems.
However, optimizing the pilot length alone is still insufficient to effectively improve the error performance or reliability in SPT, especially when the number of links is large or when the channel conditions change rapidly.
In principle, the pilot length reduction tends to worsen a system's capability of obtaining accurate channel information, resulting in performance degradations.

2) Pilot-free SPT:
Several existing works have investigated the performance of SPT under non-coherent settings.
For instance, in \cite{Lancho2020On} and \cite{Qi2020A}, the maximum coding rate of non-coherent Rayleigh block-fading channels with short block lengths was respectively analyzed for point-to-point single antenna and MIMO systems.
Moreover, a saddlepoint method was explored in \cite{Lancho2020Saddlepoint} to approximate the non-asymptotic bounds of SPT over non-coherent Rayleigh block-fading channels, enabling more precise evaluation of the BLER.
In addition, various pilot-free short packet designs have been explored in some recent efforts for URLLC.
Specifically,
in \cite{Li2021Constellation},
massive single-input multiple-output (SIMO) systems using short packets with a noncoherent maximum-likelihood detection scheme were studied where the optimal constellation structures were designed to maximize the diversity gain.
Unfortunately, the computational complexity of such noncoherent maximum likelihood detectors may be prohibitively high.
Furthermore, in \cite{Wu2020Pilot} and \cite{Ji2019Pilot},
a pilot-less sparse vector coding scheme was investigated where the decoding process of short packets relies on the identification of nonzero positions of the received signals rather than the actual values.
However, a major shortcoming of this scheme is its low spectral efficiency, especially when the length of the sparse vector is relatively large.
Additionally, a differential subcarrier index modulation scheme was proposed in \cite{Choi2021Generalized} for pilot-free SPT where the information bits were mapped to the index of joint row and column permutation patterns of a unitary matrix.
Nevertheless, this scheme requires a large number of active subcarriers for a single user to achieve a target frequency diversity gain, which may not be realistic for URLLC given the normally sparse traffic and scarce spectrum resource.
In order to achieve a well-balanced tradeoff between system complexity and performance in URLLC with SPT, it is necessary to employ non-coherent detection with low computational complexity.

3) Differential modulation:
Due to its simple transceiver structure and low complexity, non-coherent detection based on differential modulation has gained significant attention in the past and has been widely explored in many wireless networks.
For example, it has been shown in \cite{Hochwald2000Differential} and \cite{Hughes2000Differential} that differential modulation provides a good tradeoff between receiver complexity and performance for MIMO systems.
In addition, the performance of differentially modulated amplify-and-forward and decode-and-forward cooperative communications systems was studied in \cite{Himsoon2008Differential} and \cite{Zhao2007Differential}, while the differential modulation schemes for wireless relay networks over fast fading channels were investigated in \cite{Avendi2014Performance} and \cite{Zhu2010Differential}.
Moreover, the performance of differential modulation combined with a bit interleaved coded modulation-iterative decoding scheme was investigated in \cite{Baeza2018A} and \cite{Baeza2017Performance}.
Furthermore, some existing studies have exploited constellation design-based \cite{Baeza2019Non} and non-orthogonal multiple access-based \cite{Baeza2021Orthogonal} strategies to enable multi-user detection with differential modulation.
However, these works were carried out under the assumption of infinite blocklength, except \cite{Baeza2021Orthogonal}, which derived the BLER of differential modulation without channel coding.
Consequently, the existing research on differential modulation fails to capture the interplay between achievable rate, blocklength, and BLER in SPT, making a revisit and analysis of differential modulation in the URLLC regime imperative.

4) MC schemes:
In recent years, there has been a growing amount of research studying ultra-reliable transmission design based on MC schemes.
In \cite{Zheng2021Open}, for example, the URLLC performance of a joint MC and open-loop mode scheme under heterogeneous networks was explored.
In \cite{Zhou2019On}, the SPT performance of MC under parallel additive white Gaussian noise channels and independent quasi-static fading was evaluated.
In \cite{Chen2021Impact}, the impact of correlated quasi-static Rayleigh and Nakagami-$m$ fading channels on MC was studied.
Additionally, a framework based on MC was investigated in \cite{SheImproving} for improving available coverage of URLLC in the finite blocklength regime.
Nevertheless, all these studies have assumed that perfect channel state information is known at the receiver, while the impact of pilot-less schemes such as differential modulation on diversity gain was not investigated.

\vspace{-1.25em}
\subsection{Our Contributions}
In this paper, we investigate the benefits of jointly employing differential modulation and MC schemes in SPT.
The main contributions of this paper are summarized as follows.
\begin{itemize}
\item To the best of our knowledge, this is the first attempt to explore the potential of differential modulation for SPT in URLLC and provide a comprehensive performance analysis. In particular, we first derive the maximum achievable rate and minimum achievable BLER of point-to-point ergodic fading channels with phase-shift keying (PSK) inputs by using the information spectrum (IS) bound and dependence testing (DT) bound developed in \cite{Polyanskiy2010Channel}.
    The results show that the SPT formula with PSK inputs is analogous to the normal approximation where the maximum achievable rate can be regarded as the conventional Shannon capacity minus a penalty term due to short blocklength.
    Based on the above results, we find that the effect of differential modulation and time-varying channels can be described by an effective signal-to-noise ratio (SNR). This effective SNR is then used to evaluate the maximum achievable rate and minimum achievable BLER of differential modulation.
\item For comparison, we also evaluate the performance of the pilot-assisted scheme utilizing minimum mean square error (MMSE) channel estimators. Specifically, the pilot-assisted scheme considers two widely used digital modulation schemes, namely PSK and quadrature amplitude modulation (QAM), to respectively assess the SPT performance. Furthermore, we extend the maximum achievable rate and minimum achievable BLER performance analysis of both the pilot-assisted scheme and differential modulation to MC schemes such as selection combining (SC) and maximum-ratio combining (MRC) schemes.

\item A comprehensive reliability and latency performance comparison between the pilot-assisted scheme and differential modulation is presented under a wide range of URLLC application scenarios including different Doppler values, number of the associated BSs and blocklengths.
    It is revealed that, a) considering net data rate and power fairness, the BLER performance of differential modulation exceeds that of the pilot-assisted scheme, especially when the channel varies rapidly;
    b) thanks to no need for pilot transmission and channel estimation, differential modulation can transmit more information payload than the pilot-assisted scheme for given BLER, latency, and blocklength requirements.
\end{itemize}
\vspace{-1.05em}
\subsection{Paper Outline and Notation}
The outline of this paper is as follows. Section II describes the system model. In Section III, the maximum achievable rate and minimum achievable BLER performance of differential modulation under MC schemes is developed.
The maximum achievable rate and minimum achievable BLER performance of the pilot-assisted scheme is considered in Section IV. Numerical and Simulation results are given in Section V. Section VI concludes this paper.

\emph{Notations:} ${\left(  \cdot  \right)^ {*} }$ and $\left|  \cdot  \right|$ denote the conjugate and the absolute value of a complex number, respectively.
$\mathcal{CN}\left( {0,1} \right)$ stands for the complex Gaussian distribution with mean zero and unit variance.
$\mathbb{E} \left[ \cdot \right]$ and $\mathbb{V}\text{ar}\left[ \cdot \right]$ denote the expectation and variance operations, respectively. In addition, we use ${{{\mathbb{P}}}\left( \cdot \right)}$ to indicate the probability of a set and ${{{\mathbb{P}}}\left( {\cdot\left| \cdot \right.} \right)}$ for the conditional probability.
\section{System model and Assumptions}
We consider an uplink MC network scheme that comprises one URLLC user (e.g., robot, vehicle or virtual reality device) and $K$ associated base stations (BSs)\footnote{Since the downlink and uplink system models of MC schemes can be treated interchangeably for a single-user scenario, we have focused on the performance analysis of the uplink transmission in this paper and the URLLC performance for downlink can be similarly evaluated.}.
As in \cite{Zheng2021Open, Zhou2019On, Chen2021Impact, SheImproving,Avendi2013Performance, Ma2002Accurate}, both the user and BSs are assumed to be equipped with a single antenna.
Without loss of generality, we assume that the user needs to transmit a sequence of emergency information bits $\mathbf{b} = \left\{ {{b\left[ 1 \right]},{b\left[ 2 \right]},...,{b\left[ L \right]}} \right\}$ of length $L$ to the BSs.
The bit information sequence is first encoded by a rate $R_{c}=L/J$ encoder to generate a code sequence $\mathbf{c} = \left\{ {{c\left[ 1 \right]},{c\left[ 2 \right]},...,{c\left[ J \right]}} \right\}$ of length $J$ to convey the message\footnote{As in \cite{Polyanskiy2010Channel}, we assume that the optimal encoder-decoder pair has been applied to make SPT performance analysis tractable. The gap between the theoretical result and the actual performance under a specific coding scheme will be illustrated later in this paper via simulations.}.
In order to avoid the process of channel estimation, differential phase-shift keying (DPSK) modulation is further employed in our system model.
Specifically,
we transform the code bit sequence as a series of $M$-ary phase-shift keying ($M$-PSK) information symbols
and use $\mathcal{D} = \left\{ {{e^{j2\pi m/M}},m = 0,...,M - 1} \right\}$ to denote the set of $M$-PSK symbols.
Thus, the code bit sequence of length $J$ can be transformed as a block of length $N = J/{\log _2}M$ symbols. $J$ is assumed to be an integer multiple of ${\log _2}M$.
In addition, we use $d\left[ n \right]  \in \mathcal{D}$ to represent the $n$-th symbol in the block, where $n=1,...,N$.
For differential modulation,
the information is carried in the difference of the phases of two adjacent transmitted symbols.
Therefore, the $n$-th symbol transmitted by the user is given by
\begin{equation}
\begin{aligned}
s\left[ n \right] = d\left[ n \right]s\left[ {n - 1} \right],~~~n=1,...,N,
\label{equ_differential}
\end{aligned}
\end{equation}
where $s\left[ {0} \right]=1$ is the initial reference signal.
We use ${h_k}\left[ n \right]\sim \mathcal{CN}\left( {0,1} \right)$ to represent the channel gain from the user to the $k$-th associated BS, i.e., a Rayleigh fading environment is considered.
Then, the corresponding receiver signal of $s\left[ n \right]$ at the $k$-th associated BS can be expressed as
\begin{equation}
\begin{aligned}
{r_k}\left[ n \right] = \sqrt \rho {h_k}\left[ n \right]s\left[ n \right] + {w_k}\left[ n \right],~~n=0,1,...,N,
\label{equ_received}
\end{aligned}
\end{equation}
where ${w_k}\left[ n \right] \sim \mathcal{CN}\left( {0,1} \right)$ is the noise component.
Since the average power of the channel gain and the noise is normalized, the transmit power $\rho$ is equal to the average SNR.
In addition, we assume that the channels are spatially uncorrelated and temporally fluctuating. As in \cite{Avendi2013Performance}, the channel variation of the $k$-th link is modeled according to the following first-order autoregressive (AR) model:
\begin{equation}
\begin{aligned}
{h_k}\left[ n \right] = \alpha {h_k}\left[ {n - 1} \right] + \sqrt {1 - {\alpha ^2}} {e_k}\left[ n \right],
\label{AR_model}
\end{aligned}
\end{equation}
where $0 \leqslant \alpha \leqslant1$ is the channel auto-correlation parameter
and ${e_k}\left[ n \right] \sim \mathcal{CN}\left( {0,1} \right)$ is independent from symbol to symbol.
Note that $\alpha = 1$ indicates a quasi-static fading channel and $\alpha = 0$ models a completely random (from symbol to symbol) time-varying channel\footnote{
In fact, we can readily model the channel variation of practical URLLC applications by adjusting the value of $\alpha$ in (3). For instance, a larger $\alpha$ value (indicating slow varying channels or quasi-static fading channels) is suitable for modeling communications links between slow-moving robots and central controllers in factory automation scenarios.
On the other hand, a smaller $\alpha$ value (representing fast-varying channels) is more appropriate for communications links between high-speed vehicles (including UAVs) and BSs or roadside units.
}.
By using Jakes' autocorrelation model, the parameter $\alpha$ can be further written as
\begin{equation}
\begin{aligned}
\alpha  = {\mathcal{J}_0}\left( {2\pi {f_d}{T_s}} \right),
\label{AR_corr}
\end{aligned}
\end{equation}
where $\mathcal{J}_{0}\left( x  \right)$ is the zero-th order Bessel function of the first kind, $f_d$ is the Doppler frequency and $T_s$ is the symbol duration.
Based on two adjacent received symbols, each BS in MC schemes can differentially decode the signals and obtain the corresponding decision variable.
For DPSK, the decision variable of the $k$-th link can be computed as
\begin{equation}
\begin{aligned}
{z_k}\left[ n \right] =   {{r_k}\left[ n \right]r_k^ * \left[ {n - 1} \right]}.
\label{equ_diffdetection}
\end{aligned}
\end{equation}
In addition, we suppose that the short packets received by each BS in MC schemes are forwarded to the core network by backhaul links for further combining, as depicted in Fig. 1.
To obtain the diversity gain, the SC scheme and MRC scheme are respectively adopted in our system model to combine the multiple copies of the information received from the associated branches to a single resultant output.
For the SC scheme, the decision variable with the maximum magnitude is selected to recover the transmitted information,
the output is given by \cite{Avendi2014Performance}
\begin{equation}
\begin{aligned}
z\left[ n \right]  = z_{\widehat K}\left[ n \right],
\label{equ_SCcombiner}
\end{aligned}
\end{equation}
where $\widehat K = \mathop {\arg \max }\limits_{k \in \left\{ {1,...,K} \right\}} \left| {{z_k}}\left[ n \right] \right|$.
On the other hand, the MRC scheme combines all the decision variables, and the output can be expressed as \cite{Ma2002Accurate}
\begin{equation}
\begin{aligned}
z\left[ n \right]  = \sum\limits_{k = 1}^K {{z_k}\left[ n \right]} .
\label{equ_MRCcombiner}
\end{aligned}
\end{equation}
Furthermore, a soft decision detector is to be employed to process the output of the combiner $z\left[ n \right]$, attaining better estimates of the original data.
Finally,
the decoder utilizes the output of the soft decision detector to produce an estimate for the $L$-bit information: $\widehat{\mathbf{b}}=\left\{ {\widehat{b}\left[ 1 \right],\widehat{b}\left[ 2 \right],...,\widehat{b}\left[ L \right]} \right\}$.
\section{SPT with Differential Modulation}
In this section,
we first briefly review the SPT performance results of \cite{Polyanskiy2010Channel}, \cite{Polyanskiy2011Scalar} and \cite{Lancho2018Normal} for the case without considering any particular modulation scheme.
Since the phase difference between the two consecutive PSK symbols is utilized in differential modulation to transmit information, we next characterize the maximum achievable rate and minimum achievable BLER performance of a point-to-point ergodic fading channel with independent and identically distributed (i.i.d) PSK inputs and perfect CSI from an information theoretic perspective.
Based on the obtained results, the performance of SPT with MC schemes and differential modulation is further analysed by characterizing the effect of differential modulation and time-varying channels as a reduction in the effective SNR.

\subsection{Preliminaries}
In this subsection, we give a brief review of the SPT performance results
in the relevant references.
Let $\varepsilon  = \mathbb{P}\left[ {{\bf{b}} \ne {\bf{\hat b}}} \right]$ represent the probability that the receiver makes an erroneous guess about the information bit sequence $\mathbf{b}$ (i.e., BLER).
According to the results in \cite{Polyanskiy2010Channel}, \cite{Polyanskiy2011Scalar} and \cite{Lancho2018Normal},
the maximum achievable rate (bits/symbol, bits/s/Hz or bits/channel use) for point-to-point ergodic fading channels with given BLER $\varepsilon$ and blocklength $N$ can be tightly approximated as
\begin{equation}
\begin{aligned}
{\widehat R} \left( N, \varepsilon \right) \approx C\left( \rho \right) - \sqrt {\frac{V\left( \rho \right) }{N}} {Q^{ - 1}}\left( \varepsilon \right)+\frac{{{{\log }_2}N}}{{2N}},
\label{rate_fbl}
\end{aligned}
\end{equation}
where $C\left( \rho \right) = \mathbb{E}\left[{\log _2}\left( {1 + \rho {\left| h \right|^2}} \right)\right]$ is the capacity of an ergodic fading channel,
\begin{equation}
\begin{aligned}
V\left( \rho \right) = \mathbb{V}\text{ar}\left[{\log _2}\left( {1 + \rho {\left| h \right|^2}} \right)\right]+1-{\mathbb{E}^2}\left[ {\frac{1}{{1 + \rho {{\left| h \right|}^2}}}} \right]
\label{disp_contin}
\end{aligned}
\end{equation}
is termed fading channel dispersion and ${Q^{ - 1}}\left( \cdot \right)$ is the inverse of the Gaussian Q-function.
The expression (\ref{rate_fbl}), which is generally referred to as normal approximation, relies on a central-limit-theorem analysis and characterizes the asymptotic behavior of the non-asymptotic upper and lower bounds of $\widehat R$.
Several references, such as \cite{Johan2019Short,Ferrante2018Pilot,Polyanskiy2010Channel,Johan2021URLLC,Lancho2023Cell,Kislal2023Efficient,Lancho2020On,Qi2020A,Lancho2020Saddlepoint} and \cite{Polyanskiy2011Scalar, Lancho2018Normal, Durisi2016Short, A2014Jazi}, have also proven that the SPT performance for the channel models of interest in wireless communications systems can be characterized by the means of non-asymptotic bounds.
In addition, for given information bit length $L$, the minimum achievable BLER $\widehat {\varepsilon  }$ under point-to-point ergodic fading channels and blocklength $N$ can be equivalently written as [2, Eq.(23)]
\begin{equation}
\begin{aligned}
\widehat \varepsilon  \approx Q\left( {\sqrt{\frac{N}{V\left( \rho \right)}}\left( {C\left( \rho \right) - R}+\frac{{{{\log }_2}N}}{{2N}}\right)} \right),
\label{bler_fbl}
\end{aligned}
\end{equation}
where $R = \frac{L}{N}$ (bits/symbol).

\newcounter{TempEqCnt}                        
\setcounter{TempEqCnt}{\value{equation}} 
\setcounter{equation}{14}
\begin{figure*}[hb]
\hrulefill
\begin{equation}
\begin{aligned}
&I_{\text{coh}}\left( \rho \right) =\mathbb{E}\left[ {i\left( {X;Y} \right)} \right]
\mathop  = \limits^{(a)}{\log _2}M - \\
&\frac{1}{M}\sum\limits_{j = 1}^M {\mathbb{E}\left[ {{\log _2}\frac{{\sum\limits_{i = 1}^M \exp \left[ { - {{{{\left| {y - \sqrt{ \rho} h x_i} \right|}^2}}}} \right] }}{\exp \left[ { - {{{{\left| {y - \sqrt{ \rho} h x_j} \right|}^2}}}} \right]}} \right]}\mathop  = \limits^{(b)}{\log _2}M -\frac{1}{M}\sum\limits_{i = 1}^M {\mathbb{E}}\left[ {\log _2}\sum\limits_{i = 1}^M {\exp \left( {{w^2} - {{\left( {w + \sqrt \rho  h\left( {{x_j} - {x_i}} \right)} \right)}^2}} \right)}  \right]
\label{mutu_info_coh}
\end{aligned}
\end{equation}
\end{figure*}
\setcounter{equation}{\value{TempEqCnt}}

\newcounter{TempEqCnt1}                        
\setcounter{TempEqCnt1}{\value{equation}} 
\setcounter{equation}{15}
\begin{figure*}[hb]
\hrulefill
\begin{equation}
\begin{aligned}
&V_{\text{coh}} \left( \rho \right) =\mathbb{V}\text{ar}\left[ {i\left( {X;Y} \right)} \right]\mathop = \limits^{(c)}\mathbb{V}\text{ar}\left[  {\log _2}\frac{{\sum\nolimits_{i = 1}^M {{{\mathbb{P}}_{Y\left| X \right.}}\left( {y\left| x_i \right.} \right)} }}{{{{\mathbb{P}}_{Y\left| X \right.}}\left( {y\left| x \right.} \right)}}    \right]\mathop  = \limits^{(d)}\\
&\frac{1}{M}\sum\limits_{j = 1}^M {\mathbb{E}}\left[ \left\{ { {\log _2}\sum\limits_{i = 1}^M {\exp \left( {{w^2} - {{\left( {w + \sqrt \rho  h\left( {{x_j} - {x_i}} \right)} \right)}^2}} \right)} } \right\}^2  \right]-\left\{\frac{1}{M}\sum\limits_{j = 1}^M {\mathbb{E}}\left[ { {\log _2}\sum\limits_{i = 1}^M {\exp \left( {{w^2} - {{\left( {w + \sqrt \rho  h\left( {{x_j} - {x_i}} \right)} \right)}^2}} \right)} } \right]\right\}^2
\label{disp_coh}
\end{aligned}
\end{equation}
\end{figure*}
\setcounter{equation}{\value{TempEqCnt1}}

Note that the normal approximation results developed in \cite{Polyanskiy2010Channel}, \cite{Polyanskiy2011Scalar} and \cite{Lancho2018Normal} focus on the analysis of ideal continuous signals.
Since specific discrete modulation schemes are employed in practice to transmit information over wireless channels, the channel input is no longer continuous values, but rather constrained to be from a discrete constellation whose size and values are determined by the modulation type and order.
However, the normal approximation analysis for SPT with discrete inputs is not trivial.
In order to characterize the performance of SPT with discrete inputs, we resort to lower and upper bounds on BLER that are respectively based on the information spectrum (IS) bound [10, Th. 11] and the dependence testing (DT) bound [10, Th. 17].
For the sake of completeness, the IS and DT bounds proposed by \cite{Polyanskiy2010Channel} are described in the following Theorems 1 and 2 which are later used in our analysis.
Before stating these bounds, we need to define some notations.
Let $X \in \mathcal{X}$ and $Y\in \mathcal{Y}$ respectively denote the input and output symbol of the channel, where $\mathcal{X}$ is the input alphabet and $\mathcal{Y}$ is the output alphabet.
Then, the channel can be modeled with a conditional transition probability
${{\mathbb{P}}_{Y^{N}\left| X^{N} \right.}}\left( {y^{N}\left| x^{N} \right.} \right) :\mathcal{X}^{N}  \to \mathcal{Y}^{N}$ for blocklength $N$.
In addition, we define
\begin{equation}
\begin{aligned}
i\left( {X^{N};Y^{N}} \right) = \log_2 \frac{{{\mathbb{P}}_{Y^{N}\left| X^{N} \right.}}\left( {y^{N}\left| x^{N} \right.} \right)}{{{\mathbb{P}}_{Y^{N}}}\left( y^{N} \right)}
\label{Th1_inf_dens}
\end{aligned}
\end{equation}
as the information density. Then, the IS and DT bounds can be respectively described as
\newtheorem{theorem}{Theorem}
\begin{theorem}
(IS lower bound [10, Th. 11]): For a general point-to-point channel ${{\mathbb{P}}_{Y^{N}\left| X^{N} \right.}}\left( {y^{N}\left| x^{N} \right.} \right)$, the BLER $\varepsilon$ with information bits of length $L$ satisfies
\begin{equation}
\begin{aligned}
\varepsilon  \geqslant \mathop {\sup }\limits_{\beta  > 0} \left\{ {\mathop {\inf }\limits_{{{\mathbb{P}}_X}} {\mathbb{P}}\left[ {i\left( {X^{N};Y^{N}} \right) \leqslant {{\log }_2}\beta } \right] - \frac{\beta }{{{2^L}}}} \right\}
\label{Th0_lower}
\end{aligned}
\end{equation}
where
$\beta>0$ is an arbitrary positive constant.
\end{theorem}

\newtheorem{theorem0}{Theorem}
\begin{theorem}
(DT upper bound [10, Th. 17]): For a general point-to-point channel ${{\mathbb{P}}_{Y^{N}\left| X^{N} \right.}}\left( {y^{N}\left| x^{N} \right.} \right)$, the BLER $\varepsilon$ with information bits of length $L$ satisfies
\begin{equation}
\begin{aligned}
\varepsilon  \leqslant \mathbb{E}\left[ {\exp \left\{ { - {{\left[ {i\left( {{X^N};{Y^N}} \right) - {{\log }_2}\left( {\left({{{2^{L}} - 1}}\right)/{2}} \right)} \right]}^ + }} \right\}} \right],
\label{Th1_DT}
\end{aligned}
\end{equation}
where ${\left[ A \right]^ + }$ = $A$ if $A \geqslant 0$ and 0 otherwise.
\end{theorem}

Clearly, the IS bound demonstrates an impossibility result:
for a given blocklength, no encoding and decoding schemes that lie below the IS bound can be found.
In contrast, the DT bound indicates the existence of an encoding and decoding scheme (although the DT bound does not demonstrates how to construct such an encoding and decoding scheme explicitly).
In information theory, the IS and DT bounds are generally referred to as converse and achievability bounds on the error probability, respectively \cite{Cover2006Elements}.
However, the IS and DT bounds require time-consuming Monte-Carlo simulations to obtain the corresponding BLER results.
Therefore, Theorem 1 and 2 do not reflect the effect of system parameters (e.g., modulation order, SNR, blocklength and coding rate) on the performance of SPT from an engineering point of view.
To obtain an analytical BLER performance result, we need to derive the expectation and variance of the information density under coherent ergodic fading channels for a single channel use. Both are key parameters to determine the performances of SPT, as shown next in Theorem 3.
\subsection{SPT with i.i.d PSK inputs and perfect CSI over point-to-point ergodic fading channels }
In this subsection, we first derive the expectation and variance of the information density $i\left( {X;Y} \right) = \log_2 \frac{{{\mathbb{P}}_{Y\left| X \right.}}\left( {y\left| x \right.} \right)}{{{\mathbb{P}}_{Y}}\left( y \right)}$  under coherent fading channels for a single channel use with PSK inputs.
Based on the expectation and variance of the information density, the performance of SPT under a point-to-point fading channel with i.i.d PSK inputs and perfect CSI is then analysed by using the IS and DT bounds.


It is well known that the conditional transition probability for PSK modulated fading channels with unit noise variance and soft decision is given by
\begin{equation}
\begin{aligned}
{ \mathbb{ P}^{\text{psk}}_{Y\left| X \right.}}\left( {{y}\left| {{x}} \right.} \right) = \frac{1}{{ {\pi } }}\exp \left[ { - {{{{\left| {y - \sqrt{ \rho} h x} \right|}^2}}}} \right],
\label{channel_trans_proba}
\end{aligned}
\end{equation}
where the input is constrained to the $M$-ary PSK constellation (i.e., $x \in \mathcal{D}$).
According to the definition of the mutual information in (\ref{Th1_inf_dens}), we have
${i\left( {X;Y} \right)}={\log _2}M - {\log _2}\frac{{\sum\nolimits_{i = 1}^M {{{\mathbb{P}}_{Y\left| X \right.}}\left( {y\left| x_i \right.} \right)} }}{{{{\mathbb{P}}_{Y\left| X \right.}}\left( {y\left| x \right.} \right)}}$ for a single channel use, where $x_i$ denotes the $i$-th constellation point in the set $\mathcal{D}$.
Then, the expectation and variance of ${i\left( {X;Y} \right)}$ for fading channels with i.i.d PSK inputs can be respectively derived as in (\ref{mutu_info_coh}) and (\ref{disp_coh}), where $(a)$ follows from (\ref{channel_trans_proba}), $(b)$ is obtained by using ${y - \sqrt{ \rho} h x_j}=w$ and $w\sim \mathcal{CN}\left( {0,1} \right)$, $(c)$ follows from $\mathbb{V}\text{ar}\left( {a - X} \right) = \mathbb{V}\text{ar}\left( X \right)$ for any constant $a$ and random variable $X$, and  $(d)$ is due to $\mathbb{V}\text{ar}(X) = {\mathbb{E}}({X^2}) - {{\mathbb{E}}^2}(X)$.
In addition, the expectation operator can be realized by averaging the items in (\ref{mutu_info_coh}) and (\ref{disp_coh}) over the distribution of $h$ and $w$.
We further note that the expectation of the information density in (\ref{mutu_info_coh}) is the ergodic fading channel capacity with i.i.d PSK inputs.
To characterize the performances of SPT with i.i.d PSK inputs, we let $\widehat {L} \triangleq  \sup \left\{ {L:\exists \frac{L}{N} \leqslant {\widehat R}\left( {N,\varepsilon } \right)} \right\}$ denote the largest number of bits that can be transmitted for given blocklength $N$ and BLER $\varepsilon$.
Then, based on the results of (\ref{mutu_info_coh}) and (\ref{disp_coh}), we have the following theorem for the SPT performance of point-to-point ergodic fading channels with PSK inputs and perfect CSI.

\newtheorem{theorem1}{Theorem}
\begin{theorem}
(PSK inputs with perfect CSI):
For point-to-point i.i.d ergodic fading channels with i.i.d PSK channel inputs and perfect CSI, the maximum achievable rate can be determined by employing the IS and DT bounds as
\setcounter{equation}{16}
\begin{equation}
\begin{aligned}
&{\widehat R}\left( {N,\varepsilon } \right)= \frac{{{\widehat L }}}{N}  \approx  I_{\text{coh}}\left( \rho \right)- {{\sqrt {\frac{ {V_{\text{coh}}\left( \rho \right)}  }{N}}   }}Q^{-1}\left( \varepsilon \right)+\frac{{{{\log }_2}N}}{{2N}}.
\label{rate_fbl_coh}
\end{aligned}
\end{equation}
The minimum achievable BLER for any given rate $R$ and blocklength $N$ can be approximately expressed as
\begin{equation}
\begin{aligned}
\widehat \varepsilon  \approx  Q\left( {\sqrt{\frac{N}{ V_{\text{coh}} \left( \rho \right)}}\left( { I_{\text{coh}} \left( \rho \right) - R}+\frac{{{{\log }_2}N}}{{2N}}\right)} \right).
\label{bler_fbl_coh}
\end{aligned}
\end{equation}
\end{theorem}

\emph{Proof:}
See Appendix A.
$\hfill\blacksquare$

We note that, the results in Theorem 3 are basically analogous to the normal approximation (\ref{rate_fbl}) and (\ref{bler_fbl}) for the optimal continuous input fading channel, with capacity and dispersion replaced by ${I_{\text{coh}} \left( \rho \right)}$ and ${V_{\text{coh}} \left( \rho \right)}$.
Thus, ${I_{\text{coh}} \left( \rho \right)}$ and ${V_{\text{coh}} \left( \rho \right)}$ can be respectively regarded as the ergodic fading channel capacity and dispersion with i.i.d PSK inputs and perfect CSI.
\subsection{SPT with differential modulation and MC schemes over time-varying channels}

Based on the results in the Theorem 3, we in this subsection will analyse the maximum achievable rate and minimum achievable BLER performance of time-varying fading channels with differential modulation and MC schemes.
To this end, we give an equivalent representation of differential modulation, which allows the results of Theorem 3 to be used.
Then, we describe the effect of differential modulation and channel time-variation by a loss in the effective SNR.

To extend the performance results of SPT in Theorem 3 to differential modulation and MC schemes, we let ${\theta_k} = {{\left| {{h_k}} \right|}^2} $ for the $k$-th link. Then, we have ${\tilde \theta _K} = \mathop {\max } \left\{ {\theta_1},{\theta_2},...,{\theta_K} \right\}$ for the SC scheme, and ${\tilde \theta _K} =\sum\limits_{k = 1}^K  {\theta_k} $ for the MRC scheme.
Since ${\theta_k}$ has an exponential distribution in a Rayleigh fading environment, the probability density function (PDF) and cumulative distribution function (CDF) of ${\theta_k}$
can be respectively expressed as
${f_{ {\theta_k}}}\left( \theta   \right) =  \exp \left( { - {\theta } } \right)$ and ${F_{{\theta_k}}}\left( \theta  \right) = 1 - \exp \left( { - \theta} \right)$.
In addition,
the PDF of $\tilde \theta _{K}$ can be written as
${f_{\tilde \theta _{K}}}\left( \theta \right) = K {\left( {1 - \exp \left( { - \theta } \right)} \right)^{K - 1}}\exp \left( { - \theta } \right)$ for the SC scheme and ${f_{\tilde \theta _{K}}}\left( \theta \right) = \frac{{{\theta ^{K - 1}}}}{{\left( {K - 1} \right)!}}\exp \left( {-\theta} \right)$ for the MRC scheme.

\newcounter{TempEqCnt2}                        
\setcounter{TempEqCnt2}{\value{equation}} 
\setcounter{equation}{28}
\begin{figure*}[hb]
\hrulefill
\begin{equation}
\begin{aligned}
{r}_{k}\left[ n_p+\upsilon \right]& = \sqrt \rho\left( \widehat \alpha \left( \upsilon  \right) {h}_{k}\left[ {n_p} \right] + \sqrt {1 - \left({{\widehat \alpha }}\left( \upsilon  \right)\right)^2} {e}_{k}\left[ n_p +\upsilon\right] \right)s\left[ n_p+\upsilon \right]+ {w}_{k}\left[ n_p +\upsilon\right]\\
&\mathop  = \limits^{(a)} \sqrt \rho \bigg(  \widehat \alpha \left( \upsilon  \right) \widehat h_{k}\left[ n_p \right]+ {\widehat \alpha \left( \upsilon  \right)}\widetilde h_{k}\left[ n_p \right]+\sqrt {1 - \left({{\widehat \alpha }}\left( \upsilon  \right)\right)^2} {e}_{k}\left[ n_p +\upsilon\right] \bigg) s\left[ n_p+\upsilon \right]+ {w}_{k}\left[ n_p +\upsilon\right]
\label{lemma2_MMSE}
\end{aligned}
\end{equation}
\end{figure*}
\setcounter{equation}{\value{TempEqCnt2}}

Since differential modulation requires two consecutive symbols to transmit information, we use the vector $\Phi \left[ n \right] = {\left[ {1,d\left[ n \right]} \right]^T}$ to equivalently represent the symbol $d\left[ n \right]$.
In order to transmit $\Phi \left[ n \right]$, differential modulation rotates the vector by multiplying by the symbol $s\left[ n-1 \right]$ instead of directly sending ${\left[ {1,d\left[ n \right]} \right]^T}$.
Let $\widehat \Phi  \left[ n \right] = {\left[ {s\left[ {n - 1} \right],s\left[ {n - 1} \right]d\left[ n \right]} \right]^T}$ denote the rotated $\Phi \left[ n \right]$.
Note that the first symbol in $\widehat\Phi \left[ n \right]$ equals the symbol previously sent, only the second symbol in $\widehat \Phi \left[ n \right]$ needs to be sent over the channel.
In addition, due to the fact that
$d\left[ {n - 1} \right]d\left[ n \right] \in \mathcal{D}$ for any $d\left[ {n - 1} \right]\in \mathcal{D}$ and $d\left[ n \right] \in \mathcal{D}$, we have $s\left[ {n - 1} \right]\in \mathcal{D}$ and $s\left[ {n - 1} \right]d\left[ n \right]\in \mathcal{D}$.
Thus, we can equivalently represent differential modulation of the symbol $d\left[ n \right]$ as $\widehat\Phi \left[ n \right]$ and use the results in the Theorem 3 to analyse the performance of SPT with channel inputs $\widehat\Phi \left[ n \right]$.
Fig.2 illustrates the equivalent representation of differential modulation.
Then, we can describe the effect of differential modulation and channel time-variation by a loss in the effective SNR, which will be shown in the following lemma.

\begin{figure}[!htp]
\centering
\includegraphics [width=0.4\textwidth]{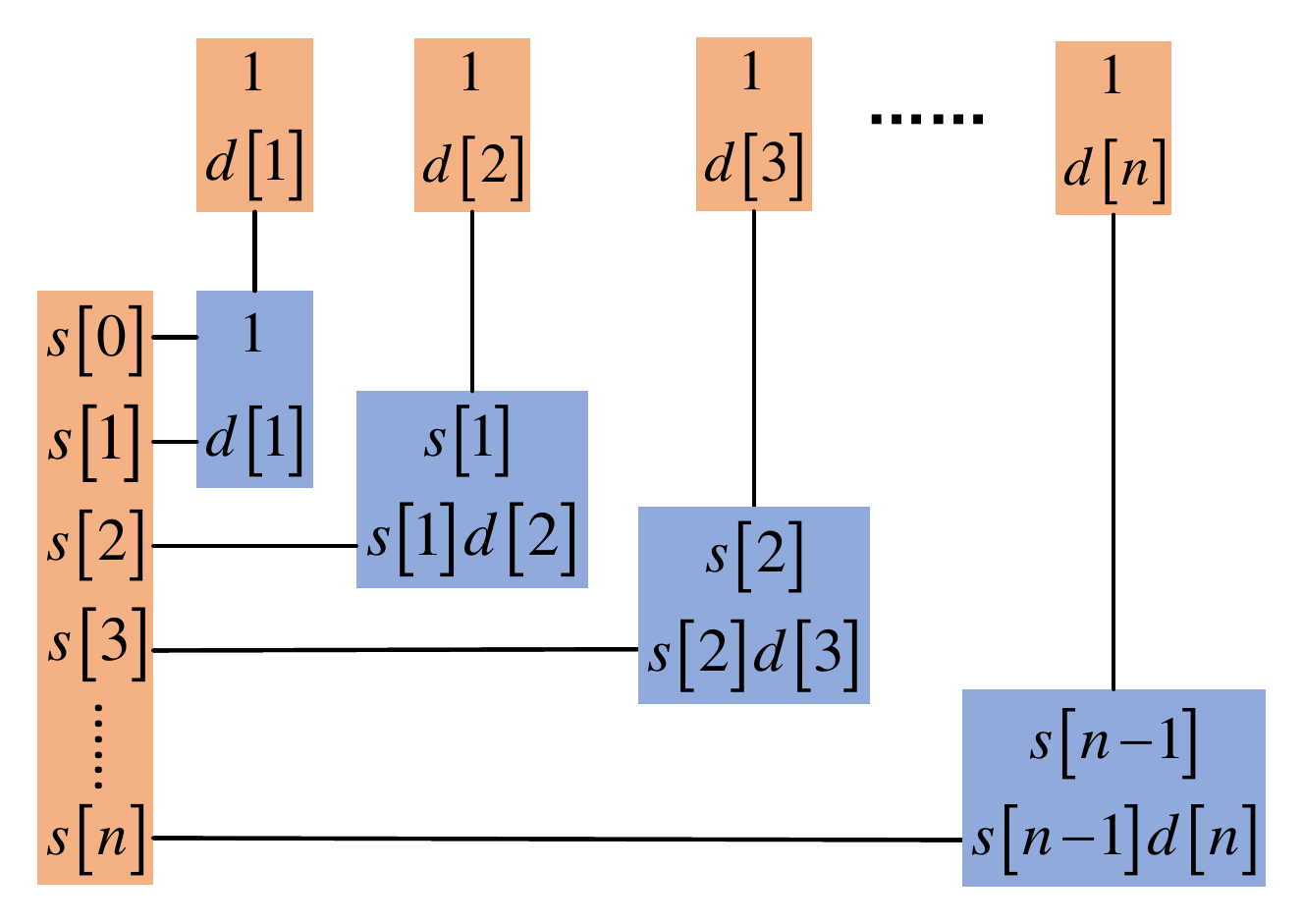}
\caption{Equivalent representation of differential modulation.}%
\label{DPSK_model}
\end{figure}

\newtheorem{lemma}{Lemma}
\begin{lemma}
\newtheorem*{Proof2}{proof}
(Effective SNR for differential modulation over time-varying channels):
With MC schemes, the effective SNR of differential modulation over time-varying channels can be quantified by
\begin{equation}
\begin{aligned}
{\rho _{\text{dif}}}\left( \tilde \theta_K\right)= \frac{{ {\rho}{\alpha ^2} \tilde \theta_K}}{{\left( {1 - {\alpha ^2}} \right){\rho} + \left( {1 + {\alpha ^2}} \right)}}.
\label{ESNR_diff_MC}
\end{aligned}
\end{equation}
\end{lemma}

\emph{Proof:}
For the channel input $\widehat\Phi \left[ n \right]$, we have
\begin{subnumcases}{}
{{r_k}\left[ {n - 1} \right] = \sqrt \rho  {h_k}\left[ {n - 1} \right]s\left[ {n - 1} \right] + {w_k}\left[ {n - 1} \right]}~~~~~~\\
{{r_k}\left[ n \right] = \sqrt \rho  {h_k}\left[ n \right]s\left[ {n - 1} \right]d\left[ n \right] + {w_k}\left[ n \right]}.
\end{subnumcases}
By substituting the first-order AR model (\ref{AR_model}) into (20b), we can obtain
\begin{equation}
\begin{aligned}
&{r}_{k}\left[ n \right] = \sqrt \rho\left( \alpha {h}_{k}\left[ {n-1 } \right] + \sqrt {1 - {\alpha ^2}} {e}_{k}\left[ n \right] \right)s\left[ n -1\right]d\left[ n \right] \\
&+ {w}_{k}\left[ n \right].
\label{lemma_AR_rec2}
\end{aligned}
\end{equation}
It should be noted that the first symbol in $\widehat\Phi \left[ n \right]$ only serves as a reference for the second symbol and does not convey any information about $d\left[ n \right]$.
Therefore, we substitute (20a) into (\ref{lemma_AR_rec2}) and obtain
\begin{equation}
\begin{aligned}
&{r}_{k}\left[ n \right] =\alpha d\left[ n \right]\left( {r_{k}\left[ {n - 1} \right] - w_{k}\left[ {n - 1} \right]} \right)+{w}_{k}\left[ n \right]+ \\
&\sqrt {{\rho}\left( {1 - {\alpha ^2}} \right)}  {e}_{k}\left[ n \right]s\left[ n -1\right]d\left[ n\right]
\label{lemma_AR_diff}
\end{aligned}
\end{equation}
In addition, all the noise terms in (\ref{lemma_AR_diff}) can be regarded as Gaussian with variance ${{\left( {1 - {\alpha ^2}} \right){\rho} + \left( {1 + {\alpha ^2}} \right)}}$.
Thus, we can write
\begin{equation}
\begin{aligned}
&{r}_{k}\left[ n \right] =\alpha d\left[ n \right]r_{k}\left[ {n - 1} \right] + \sqrt {1 + {\alpha ^2} + \left( {1 - {\alpha ^2}} \right){\rho}} \widehat w_{k}\left[ n \right],
\label{lemma_eq1}
\end{aligned}
\end{equation}
where $\widehat w_{k}\left[ n \right]$ has the same statistical properties as $ w_{k}\left[ n \right]$.
Eq.(\ref{lemma_eq1}) shows that the signal $d\left[ n \right]$ in differential modulation can be regarded as transmitting through a channel with fading response $r_{k}\left[ {n - 1} \right]$, which is known to the receiver, and corrupted by the noise term with variance ${{\left( {1 - {\alpha ^2}} \right){\rho} + \left( {1 + {\alpha ^2}} \right)}}$.
Since $r_{k}\left[ {n - 1} \right]$ has signal strength ${\rho}{{\left| h_{k} \left[ {n - 1} \right]\right|}^2}$,
the effective SNR for differential modulation over time-varying channels is given by ${{\rho} _{\text{dif}}}= \frac{{{\alpha ^2}{{\left| h_{k} \left[ {n - 1} \right]\right|}^2}{\rho}}}{{\left( {1 - {\alpha ^2}} \right){\rho} + \left( {1 + {\alpha ^2}} \right)}}$.
Finally, the effective SNR results in (\ref{ESNR_diff_MC}) are established by replacing the channel power gain with $\tilde \theta_K$(due to the MC scheme). $\hfill\blacksquare$

It should be noted that the effective SNR in (\ref{ESNR_diff_MC}) for a quasi-static fading channel is $\mathop {\lim }\limits_{\alpha \to 1} {{ \rho} _{\text{dif}}} =\frac{\rho {\tilde \theta_K} }{2}$.
This factor of 2 is the well-known 3 dB loss in performance when employing differential modulation versus coherent modulation schemes.
In fact, the 3-dB penalty that DPSK suffers in comparison with PSK is due to the effective doubling of the additive noise power for $\widehat\Phi \left[ n \right]$.
Based on Lemma 1, we can obtain the mutual information $ I_{\text{dif}} \left( \rho \right)$ and channel dispersion $ V_{\text{dif}} \left( \rho \right)$ of differential modulation and MC schemes
by substituting (\ref{ESNR_diff_MC}) into (\ref{mutu_info_coh}) and (\ref{disp_coh}). That is to say, $ I_{\text{dif}} \left( \rho \right) =I_{\text{coh}} \left( \rho_{\text{dif}} \right)$ and $ V_{\text{dif}} \left( \rho \right) =V_{\text{coh}} \left( \rho_{\text{dif}} \right)$. We further note that the time-variation of the channel has an impact on $ I_{\text{dif}} \left( \rho \right)$ and $ V_{\text{dif}} \left( \rho \right)$ of differential modulation.
The reason is that the phase difference between consecutive symbols is exploited in differential modulation to transmit data.
In contrast, the mutual information of PSK transmission with perfect CSI is independent of the channel variation.
Then, the performance of SPT with differential modulation and MC schemes can be obtained by replacing the channel capacity and dispersion in Theorem 2 with $ I_{\text{dif}} \left( \rho \right)$ and $ V_{\text{dif}} \left( \rho \right)$.
Finally, we have the following corollary for differential modulation over fading channels.

\newtheorem{corollary}{Corollary}
\begin{corollary}
\newtheorem*{Proof4}{proof}
(Differential modulation).
The maximum achievable rate of differential modulation over time-varying fading channels can be determined as
\begin{equation}
\begin{aligned}
&{\widehat R}\left( {N,\varepsilon } \right)= \frac{{{\widehat L }}}{N}  \approx  I_{\text{dif}}\left( \rho \right)- {{\sqrt {\frac{ {V_{\text{dif}}\left( \rho \right)}  }{N}}   }}Q^{-1}\left( \varepsilon \right)+\frac{{{{\log }_2}N}}{{2N}}.
\label{rate_fbl_diff_fading}
\end{aligned}
\end{equation}
The minimum achievable BLER over time-varying fading channels can be approximately expressed as
\begin{equation}
\begin{aligned}
\widehat \varepsilon  \approx  Q\left( {\sqrt{\frac{N}{ V_{\text{dif}} \left( \rho \right)}}\left( { I_{\text{dif}} \left( \rho \right) - R}+\frac{{{{\log }_2}N}}{{2N}}\right)} \right).
\label{bler_fbl_diff_fading}
\end{aligned}
\end{equation}
\end{corollary}

\section{SPT with pilot-assisted schemes}
As mentioned in Section I, the SPT performance analysis often assume that the receiver either knows the channel, or acquires their estimates by known pilot symbols.
In this section, the maximum achievable rate and minimum achievable BLER of the pilot-assisted scheme are analysed for comparison.
As in \cite{Mousaei2017Optimizing, Cao2022Independent, Ren2020Joint}, we assume that $N_p$ pilot symbols are periodically inserted in data.
In addition, the MMSE estimator is employed to obtain estimated channel information for each pilot symbol.
Specifically, the MMSE channel estimate of a known pilot symbol in $k$-th link can be obtained as \cite[Eq.(7)]{Hassibi2003How}
\begin{equation}
\begin{aligned}
\widehat h_k\left[ n_p \right] =\frac{1}{{\sqrt \rho  }}  {\left( {\frac{1}{\rho } + {{\left| {s\left[ {{n_p}} \right]} \right|}^2}} \right)^{ - 1}}{s^ * }\left[ {{n_p}} \right]{r_k}\left[ {{n_p}} \right]
\label{equ_ml}
\end{aligned}
\end{equation}
for $n_p=1,...,N_p$, where ${s }\left[ n_p \right]$ and $r_k\left[ n_p \right]$ are the channel input (i.e. the pilot) and output in the $k$-th link.
Moreover, we use $\widetilde h_k\left[ n_p \right]= h_k\left[ n_p \right]-\widehat h_k\left[ n_p \right]$ to denote the channel estimation error for the MMSE estimator.
It should be noted that the procedure of (\ref{equ_differential}) and (\ref{equ_diffdetection}) is skipped in the pilot-assisted scheme. Namely, $s\left[ n \right] = d\left[ n \right]$ and ${z_k}\left[ n \right] =   {r_k}\left[ n \right]$ for all data and pilot symbols.
Furthermore, the receiver with the MMSE estimator assumes the channel during data transmission to be the same as that during pilot transmission and the estimated channel information is then utilized to detect several subsequent symbols until another pilot symbol occurs.
Owing to the often time-varying nature of the channel,
the actual channel may deviate progressively from the channel estimate obtained at time $n_p$.
The performance will be dominated by the worst channel estimation error.
Thus, we consider the effective SNR of the data symbol ${r_k}\left[ n_p+\upsilon \right]$, where  $\upsilon = N/N_{p}$ and ${r_k}\left[ n_p+\upsilon+1 \right]={r_k}\left[ n_{p+1} \right]$ as another received pilot symbol.
As before, let $\widehat \theta_{k}$ denote ${{\left|  \widehat h_k\left[ n_p \right]  \right|}^2} $ of the $k$-th link and $\overline{ \theta }_{K}$ denote $ \mathop {\max } \left\{ \widehat \theta_{1},\widehat\theta_{2},...,\widehat\theta_{K} \right\}$ of the SC scheme or $\sum\limits_{k = 1}^K  \widehat\theta_k $ of the MRC scheme for MMSE channel estimators.
The lemma below shows that the impact of time varying channels and MMSE channel estimators on the effective SNR.

\begin{lemma}
\newtheorem*{Proof3}{proof}
(Effective SNR for the pilot-assisted scheme):
With the MMSE channel estimator, the effective SNR under MC schemes is
\begin{equation}
\begin{aligned}
{\rho _{\text{MMSE}}}\left( \overline \theta_K \right)= \frac{{\left( {1 + \rho } \right)\rho {{(\hat \alpha \left( \upsilon  \right))}^2}{{\bar \theta }_K}}}{{{{\left( {1 + \rho } \right)}^2} - {{\left( {\rho \hat \alpha \left( \upsilon  \right)} \right)}^2}}}.
\label{ESNR_MMSE_MC}
\end{aligned}
\end{equation}
\end{lemma}

\emph{Proof:}
As in \cite{Peel2004Effective}, we approximate the channel gain of future $\upsilon$ symbols in the first-order AR model as
\begin{equation}
\begin{aligned}
{h_k}\left[ {n_p + \upsilon } \right] = \widehat \alpha \left( \upsilon  \right){h_k}\left[ n_p \right] + \sqrt {1 - \left({{\widehat \alpha }}\left( \upsilon  \right)\right)^2} {\widehat e_k}\left[ {n_p + \upsilon } \right],
\label{ar_future_appro}
\end{aligned}
\end{equation}
where $\widehat \alpha \left( \upsilon  \right)={\mathcal{J}_0}\left( {2\pi {f_d}{T_s}} \upsilon \right)$ and ${\widehat e_k}\left[ {n_p + \upsilon } \right] \sim \mathcal{CN}\left( {0,1} \right)$ is temporally uncorrelated.
Then, substituting (\ref{ar_future_appro}) into the channel model (\ref{equ_received}) yields (\ref{lemma2_MMSE}),
where $(a)$ is obtained by using the channel estimation error $\widetilde h_k\left[ n_p \right]= h_k\left[ n_p \right]-\widehat h_k\left[ n_p \right]$.

For the MMSE estimator, the variance of the estimation error $\widetilde h_k\left[ n_p \right]$ is given by \cite[Eq.(19)]{Hassibi2003How}: $\delta _{\widetilde h}^2 = \frac{1}{{1 + \rho }}$.
In addition, since $w_{k}\left[ n_p \right]$ and ${e}_{k}\left[ n_p +\upsilon\right]$ are equivalent in distribution,
we combine $\widetilde h_k\left[ n_p \right]$, $w_{k}\left[ n_p \right]$ and ${e}_{k}\left[ n_p +\upsilon\right]$ by adding the variances to obtain
\begin{equation*}\nonumber
\begin{aligned}
&{r}_{k}\left[ n_p+\upsilon \right] = {{\sqrt {{\rho }} }} \widehat \alpha \left( \upsilon  \right) \widehat h_{k}\left[ n_p \right]s\left[ n_p+\upsilon \right] \\
&+ \sqrt {{1+ \frac{\rho}{{1 + \rho }}\left({{\widehat \alpha }}\left( \upsilon  \right)\right)^2} + \rho \left( {1 - \left({{\widehat \alpha }}\left( \upsilon  \right)\right)^2}\right) } {\widehat w}_{k}\left[ n_p +\upsilon\right].
\end{aligned}
\end{equation*}
Finally, the effective SNR in (\ref{ESNR_MMSE_MC}) for the MMSE estimator can be established
by replacing the channel power gain with $\widehat \theta_K$(due to the MC scheme).
$\hfill\blacksquare$

Similarly to the derivation procedures in Section III for the case of differential modulation, we can therefore attain the expectation $ I^{\text{psk}}_{\text{MMSE}} \left( \rho \right)$ and variance $ V^{\text{psk}}_{\text{MMSE}} \left( \rho \right)$ of the information density for the pilot-assisted scheme with the MMSE channel estimator and PSK inputs by substituting (\ref{ESNR_MMSE_MC}) into (\ref{mutu_info_coh}) and (\ref{disp_coh}). That is to say, we have
$ I^{\text{psk}}_{\text{MMSE}} \left( \rho \right)=I_{\text{coh}} \left( \rho_{\text{MMSE}} \right)$ and $ V^{\text{psk}}_{\text{MMSE}} \left( \rho \right) =V_{\text{coh}} \left( \rho_{\text{MMSE}} \right)$.

On the other hand, the pilot percentage and length have a significant impact on the performance of the pilot-assisted scheme.
Specifically, increasing the pilot percentage can improve the channel estimation accuracy, which also implies that the available information payload or the code length in SPT is decreased.
Note that the definition of the blocklength in \cite{Durisi2016Toward} and \cite{Polyanskiy2010Channel} is the number of complex symbols after the channel encoder and the modulator. For the pilot-assisted scheme, the number of complex symbols after the channel encoder and the modulator is $\widetilde N=N-N_p$.
Then, as in \cite{Mousaei2017Optimizing,Cao2022Independent,Ren2020Joint},
the impact of the pilot percentage and length can be reflected by replacing the blocklength in (\ref{rate_fbl_coh}) and (\ref{bler_fbl_coh}) with $\widetilde N$.
Finally,
we have the following corollary for the pilot-assisted scheme with the MMSE channel estimator and PSK inputs.

\newtheorem{corollary2}{Corollary}
\begin{corollary}
\newtheorem*{Proof5}{proof}
(Pilot-assisted scheme with PSK inputs).
With the MMSE channel estimator and PSK inputs, the maximum achievable rate of time-varying fading channels can be determined as
\setcounter{equation}{29}
\begin{equation}
\begin{aligned}
{\widehat R}\left( {N,\varepsilon } \right)= \frac{{{\widehat L }}}{\widetilde N} \approx I^{\text{psk}}_{\text{MMSE}} \left( \rho \right)- {{\sqrt {\frac{ V^{\text{psk}}_{\text{MMSE}} \left( \rho \right)  }{\widetilde N}}   }}Q^{-1}\left( \varepsilon \right)+\frac{{{{\log }_2} {\widetilde N}}}{{2 {\widetilde N}}}.
\label{rate_fbl_coh_fading}
\end{aligned}
\end{equation}
The minimum achievable BLER can be approximately expressed as
\begin{equation}
\begin{aligned}
\widehat \varepsilon  \approx
Q\left( {\sqrt{\frac{\widetilde N}{V^{\text{psk}}_{\text{MMSE}} \left( \rho \right)}}\left( {I^{\text{psk}}_{\text{MMSE}} \left( \rho \right) - \frac{{{ L }}}{\widetilde N}}+\frac{{{{\log }_2}\widetilde N}}{{2\widetilde N}}\right)} \right).
\label{bler_fbl_coh_fading}
\end{aligned}
\end{equation}
\end{corollary}

\begin{figure*}[!htp]
\centering
\includegraphics [width=0.95\textwidth]{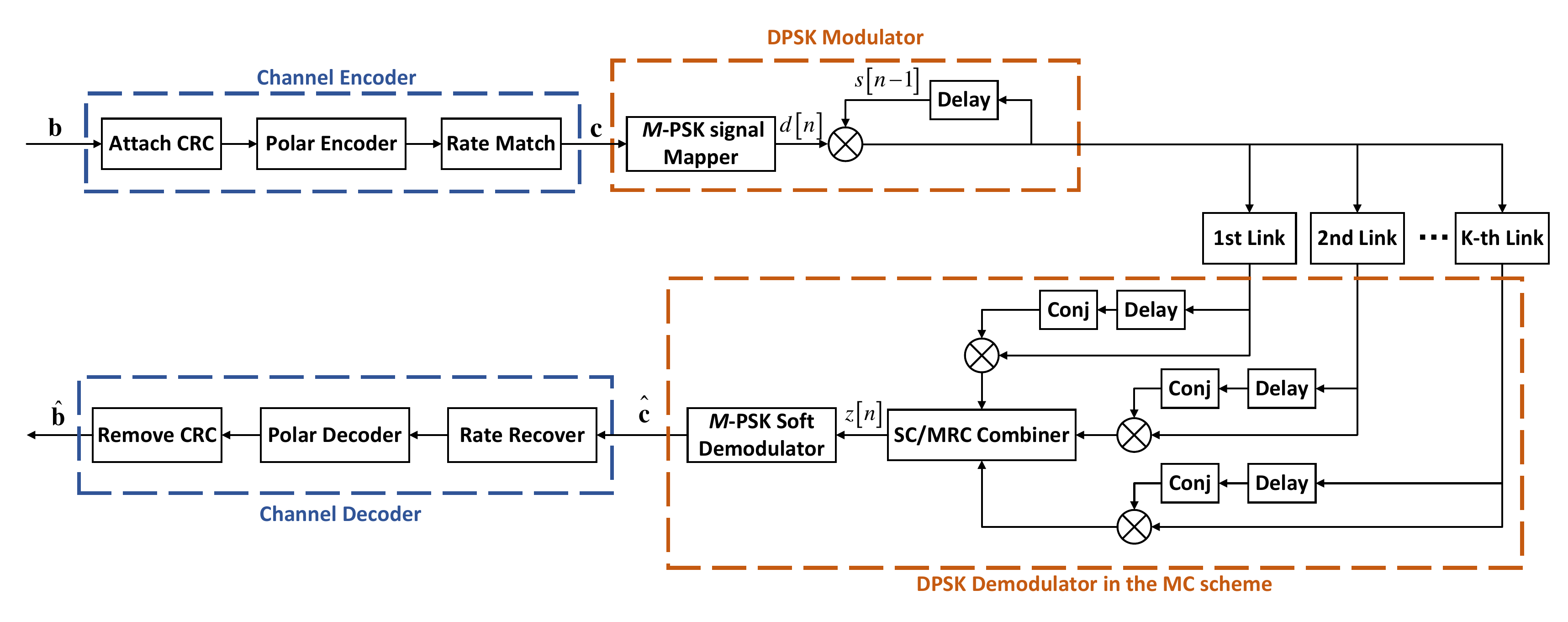}
\caption{Polar coding, differential modulation and MC schemes-based system block diagram.}%
\label{DPSKwithMC}
\end{figure*}

\begin{figure*}[!ht]
\setlength{\abovecaptionskip}{5pt}
\setlength{\belowcaptionskip}{-5pt}
\centering
\subfloat{
\includegraphics[width=0.3\textwidth]{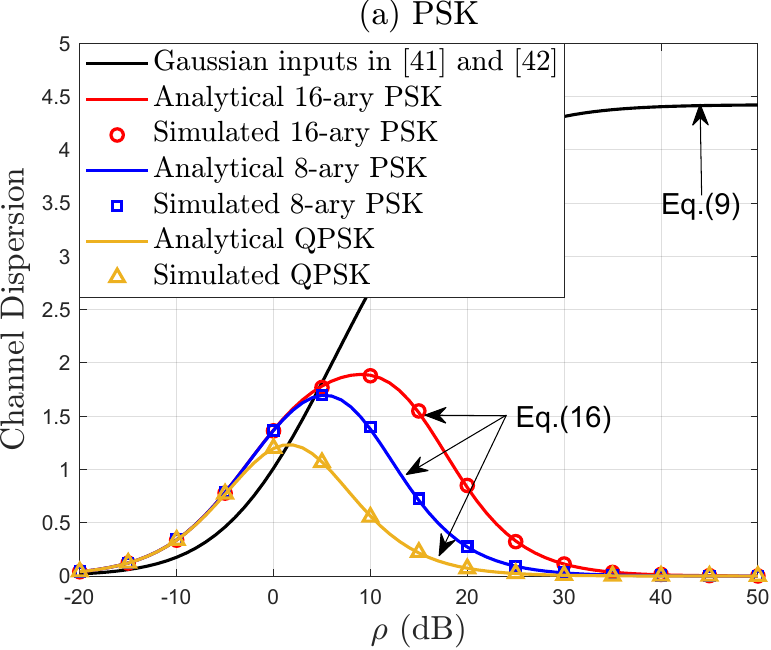}
}
\subfloat{
\includegraphics[width=0.3\textwidth]{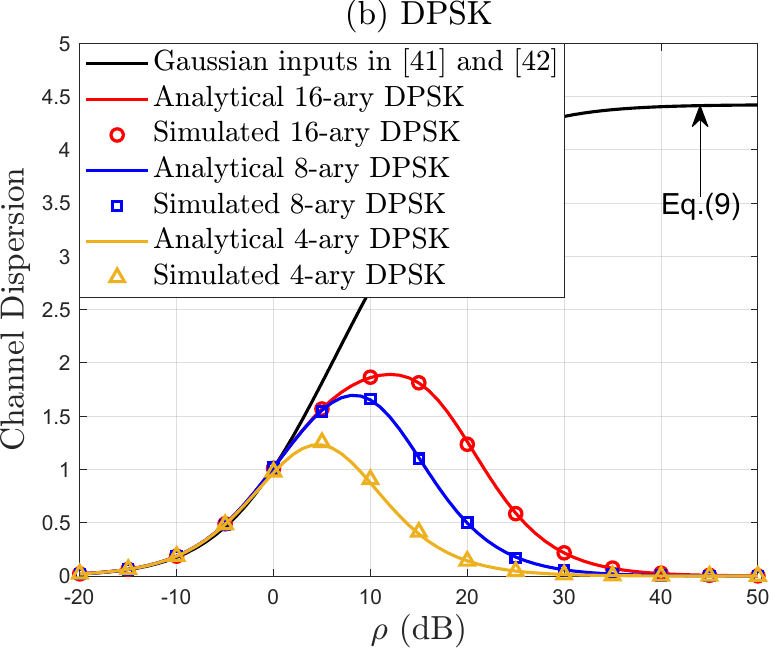}
}
\subfloat{
\includegraphics[width=0.3\textwidth]{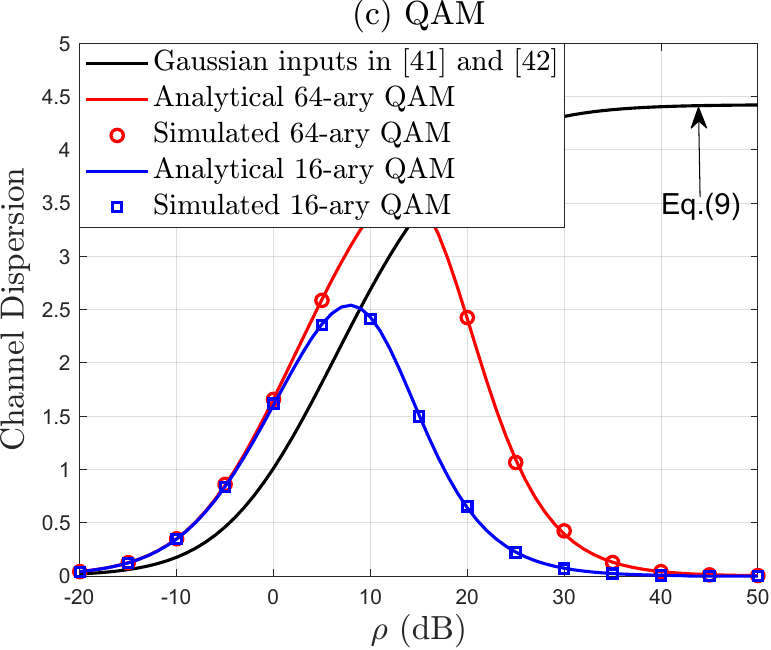}
}
\caption{The channel dispersion for fading channels with PSK, QAM, differentially modulated and Gaussian inputs.}
\end{figure*}
\setlength{\textfloatsep}{26pt}

\begin{figure*}[!ht]
\setlength{\abovecaptionskip}{5pt}
\setlength{\belowcaptionskip}{-5pt}
\centering
\begin{minipage}{1\linewidth}	
\centering
\subfloat{
\includegraphics[width=0.45\textwidth]{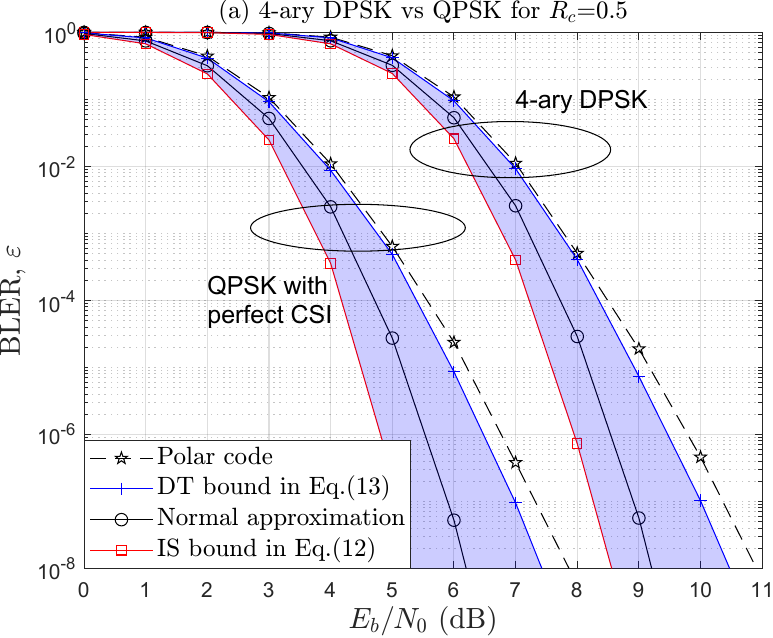}
}
\subfloat{
\includegraphics[width=0.45\textwidth]{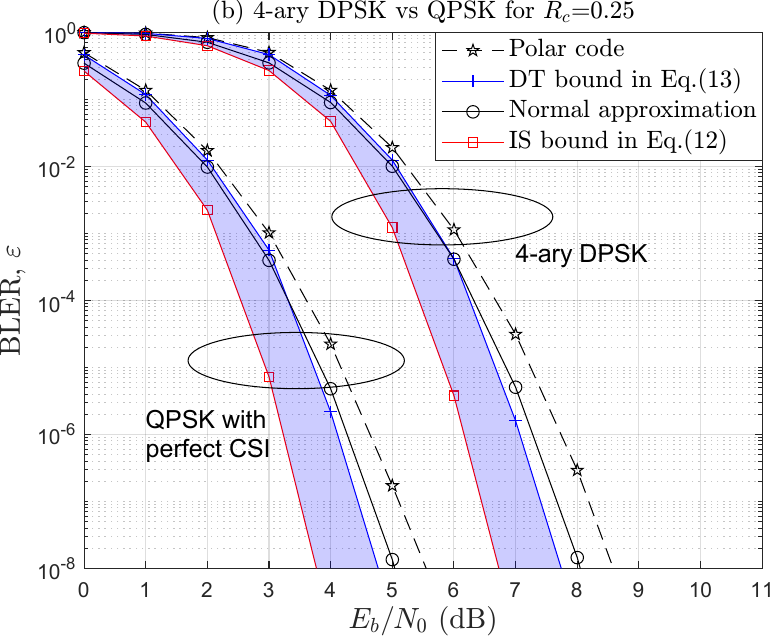}
}
\end{minipage}
\vskip 0cm 
\begin{minipage}{1\linewidth }
\centering
\subfloat{
\includegraphics[width=0.45\textwidth]{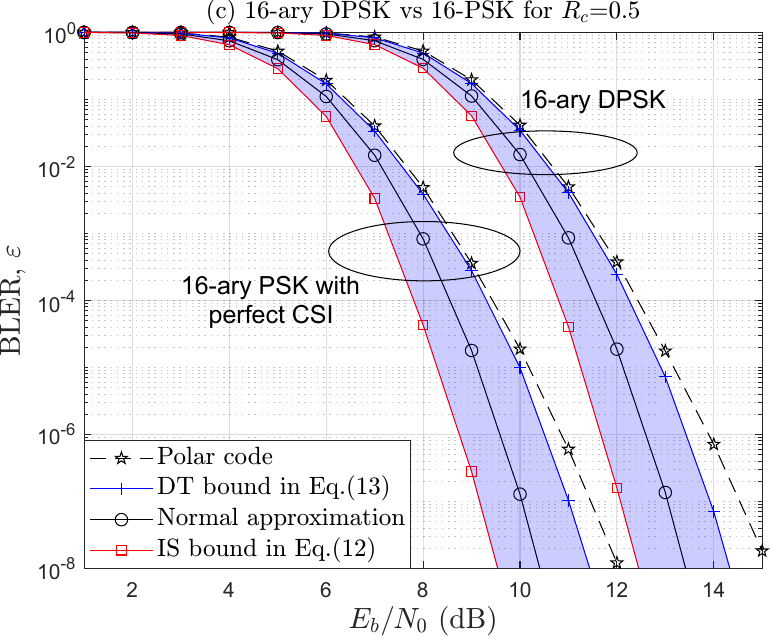}
}
\subfloat{
\includegraphics[width=0.45\textwidth]{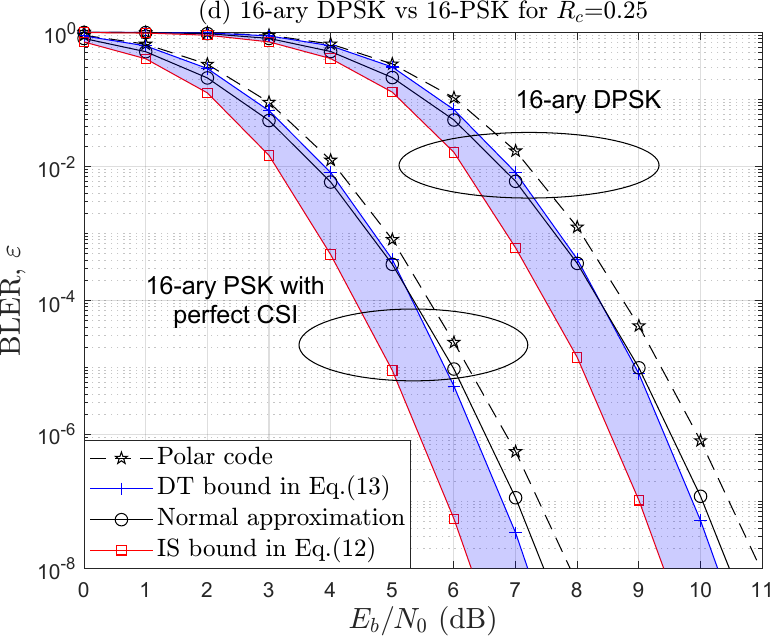}
}
\end{minipage}
\caption{BLER comparison of differential modulation and PSK with perfect CSI for different channel coding rates and modulation order values under $K=1$.}
\end{figure*}
For the pilot-assisted scheme, another common digital modulation method is QAM.
In order to obtain a comprehensive comparison, we further derive the SPT performance of the pilot-assisted scheme with QAM inputs.
Specifically, we let $\mathcal{X}^{\text{qam}} = \sqrt {\frac{1}{{{\beta }}}}  \times \left\{ { \pm \left( {2m - 1} \right) \pm \left( {2m - 1} \right)j} \right\}$ denote the input alphabet of an $M$-ary QAM constellation, where $m \in \left\{ {1,...,\frac{{\sqrt M }}{2}} \right\}$ and ${\beta } = \frac{{2\left( {M - 1} \right)}}{3}$ is a constant to normalize the average transmit power to unity.
Then, the conditional channel transition probability for QAM modulated fading channels with unit noise variance can be written as ${ \mathbb{ P}^{\text{qam}}_{Y\left| X \right.}}\left( {{y}\left| {{x}} \right.} \right) = \frac{1}{{ {\pi } }}\exp \left[ { - {{{{\left| {y - \sqrt{ \rho} h x} \right|}^2}}}} \right]$, where channel inputs $x \in \mathcal{X}^{\text{qam}}$.
It should be noted that ${ \mathbb{ P}^{\text{qam}}_{Y\left| X \right.}}\left( {{y}\left| {{x}} \right.} \right)$ is basically analogous to the conditional transition probability in (\ref{channel_trans_proba}) for PSK, with channel input constellation replaced by $\mathcal{X}^{\text{qam}}$.
Then, the channel capacity $I^{\text{qam}}_{\text{MMSE}} $ and dispersion $V^{\text{qam}}_{\text{MMSE}}$ for fading channels with QAM inputs and the MMSE estimator can be directly obtained by replacing the conditional transition probability of $I^{\text{psk}}_{\text{MMSE}} $  and $V^{\text{psk}}_{\text{MMSE}}$ with ${ \mathbb{ P}^{\text{qam}}_{Y\left| X \right.}}\left( {{y}\left| {{x}} \right.} \right)$.
Similarly, we can also have the following corollary:

\newtheorem{corollary3}{Corollary}
\begin{corollary}
\newtheorem*{Proof6}{proof}
(Pilot-assisted scheme with QAM inputs).
With the MMSE channel estimator and QAM inputs, the maximum achievable rate of time-varying fading channels can be written as
\begin{equation}
\begin{aligned}
{\widehat R}\left( {N,\varepsilon } \right)= \frac{{{\widehat L }}}{\widetilde N} \approx I^{\text{qam}}_{\text{MMSE}} \left( \rho \right)- {{\sqrt {\frac{ V^{\text{qam}}_{\text{MMSE}} \left( \rho \right)  }{\widetilde N}}   }}Q^{-1}\left( \varepsilon \right)+\frac{{{{\log }_2}\widetilde N}}{{2\widetilde N}}.
\label{rate_fbl_qam_mmse}
\end{aligned}
\end{equation}
The minimum achievable BLER can be approximately determined as
\begin{equation}
\begin{aligned}
\widehat \varepsilon  \approx  Q\left( {\sqrt{\frac{\widetilde N}{V^{\text{qam}}_{\text{MMSE}} \left( \rho \right)}}\left( {I^{\text{qam}}_{\text{MMSE}} \left( \rho \right) - \frac{{{ L }}}{\widetilde N}}+\frac{{{{\log }_2}\widetilde N}}{{2\widetilde N}}\right)} \right).
\label{bler_fbl_qam_mmse}
\end{aligned}
\end{equation}
\end{corollary}

\section{Simulation Results}

In this section, the analytical minimum achievable BLER (i.e., normal approximation) of differential modulation is explored and compared with that of the coherent modulation with the MMSE estimator.
For comparison,
Monte-Carlo simulations are further utilized to attain the simulated BLER results.
Specifically, for each SNR value, we use $10^{9}$ packet samples to respectively compute the IS bound in Theorem 1 and the DT bound in Theorem 2.
In addition, since the non-asymptotic bounds developed in \cite{Polyanskiy2010Channel} are based on optimal encoder-decoder pairs, there may be a gap between normal approximation results and the performance
achievable by the actual coding schemes. To show how close actual encoder-decoder pairs perform to the normal approximation results derived in this paper, we adopt cyclic redundancy check (CRC)-aided polar encoding and successive cancellation list (SCL) decoding schemes, as in \cite{Yuan2021Polar}.
In particular, we consider a block diagram as shown in Fig.3 to obtain the simulated BLER of SPT with polar coding, differential modulation and MC schemes.
For the coding scheme, a 6-bit CRC code with generator polynomial ${x^6} + {x^5} + 1$ is first appended at the end of $L$ information bits. Then, the CRC appended information bits are polar-encoded and rate-matched to output $J$ information bits. As a result, an ($J$, $L$) polar encoder is constructed.
For the decoding scheme of polar codes, we use an SCL decoding algorithm with a list size of 32 to generate the estimated information bits.
In polar coding-based Monte-Carlo simulations, the simulated BLER for each SNR value is calculated once 100 packet errors have been obtained.
Fig.4 provides a comparison of the fading channel dispersion with PSK, QAM, differentially modulated and Gaussian inputs.
Especially, the simulated results in Fig.4 are obtained by using the definition of the channel dispersion for a single channel use (i.e., $\mathbb{V}\text{ar}\left[ {i\left( {X;Y} \right)} \right]$).
Furthermore, since the channel variation has an influence on the channel dispersion of differential modulation, we further assume a quasi-static fading channel (i.e., $\alpha=1$) in Fig.4 to obtain the most favourable performance of differential modulation.
From Fig.4, the simulated channel dispersion results match well with the analytical results.
In addition, several interesting comparisons regarding the channel dispersion can be observed in Fig.4.
Firstly,
it can be seen that the channel dispersion with discrete inputs exhibits similar upward trends to that with optimal Gaussian inputs at low SNR.
In addition, we observe that the channel dispersion of low order modulated inputs coincides with that of high order modulated inputs for DPSK, PSK and QAM at low SNR.
Secondly, similar to the channel capacity, differential modulation suffers a 3-dB penalty in terms of the channel dispersion compared with their PSK counterpart.
Thirdly, we note that the dispersion for fading channels with optimal Gaussian inputs gradually increases with the increase of SNR and eventually reaches a maximum value.
However, the discrete modulation induced dispersion falls to zero at high SNR.
This is because, for given blocklenght $N$ and BLER $\varepsilon$, the maximum achievable rate with $M$-ary discrete inputs approaches the maximal possible $\log_2 M$ information bits at high SNR.
Therefore, unlike Gaussian inputs, short blocklength has less influence or penalty on the maximum achievable rate with discrete inputs at high SNR.



Fig.5 compares the BLER of differential modulation with that of PSK with perfect CSI for different channel coding rates and modulation order values under $K=1$.
Specifically, the transmitted packet size is 32 bytes (a common reference size for URLLC packets), resulting in (256, 128) polar codes for the coding rate $R_c=0.5$ and (256, 64) polar codes for the coding rate $R_c=0.25$.
In addition, we use $E_b$ to represent the energy per information bit and $E_b/N_0$ to denote the bit SNR. Then, we have $\rho = {{{E_b}{R_c}{{\log }_2}M}}$.
Similarly, we assume a quasi-static fading channel (i.e., $\alpha=1$) to obtain the most favourable BLER of differential modulation.
It can be seen that PSK with perfect CSI can achieve approximately a 3 dB performance advantage for the normal approximation, IS bound, DT bound and polar coding-based results when compared with differential modulation, which is consistent with their well-known BER characteristics.
Moreover, it is evident that the normal approximation results for both differential modulation and coherent PSK, with a coding rate of $R_c=0.5$, lie in a specific region (highlighted in blue) bounded by their respective IS and DT bounds.
When considering a coding rate of $R_c=0.25$, it can be observed that the normal approximation results still provide a reasonably tight approximation of the non-asymptotic bounds at low error probabilities or high SNR.
Furthermore, we note that the simulated polar coding schemes for the coding rate $R_c=0.5$ exhibit a performance gap of around 1 dB to the normal approximation at a BLER of $10^{-6}$ for both DPSK and PSK.
Consequently, the normal approximation results proposed in this paper offer satisfactory accuracy and low computational complexity in characterizing the minimum achievable BLER performance of differential modulation and PSK.

\begin{figure*}[!ht]
\setlength{\abovecaptionskip}{5pt}
\setlength{\belowcaptionskip}{-5pt}
\centering
\subfloat{
\includegraphics[width=0.45\textwidth]{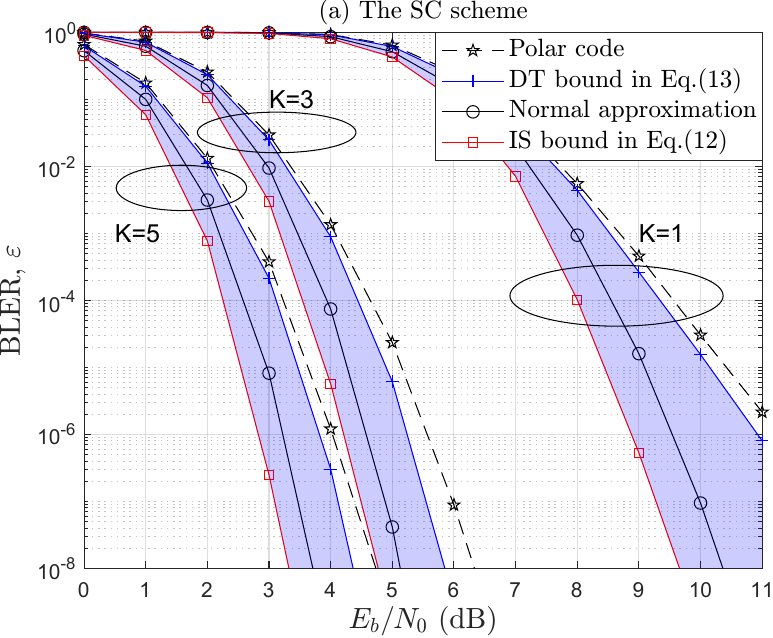}
}
\subfloat{
\includegraphics[width=0.45\textwidth]{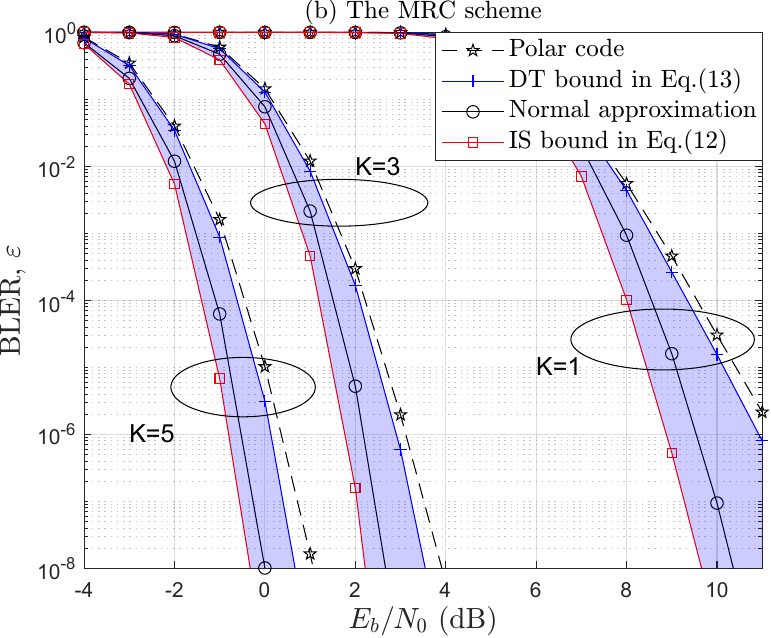}
}
\caption{The BLER performance of differential modulation for different $K$ under fading channels with the normalized Doppler value $f_dT_s = 0.05$, modulation order $M=4$ and coding rate $R_c=0.5$: (a) the SC scheme, and (b) the MRC scheme.}
\end{figure*}
\setlength{\textfloatsep}{26pt}

\begin{figure*}[!ht]
\setlength{\abovecaptionskip}{5pt}
\setlength{\belowcaptionskip}{-5pt}
\centering
\subfloat{
\includegraphics[width=0.45\textwidth]{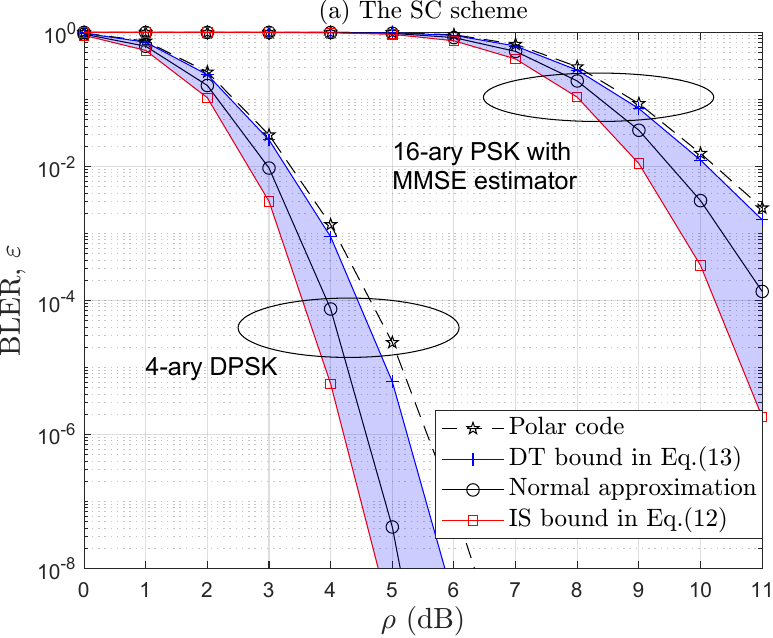}
}
\subfloat{
\includegraphics[width=0.45\textwidth]{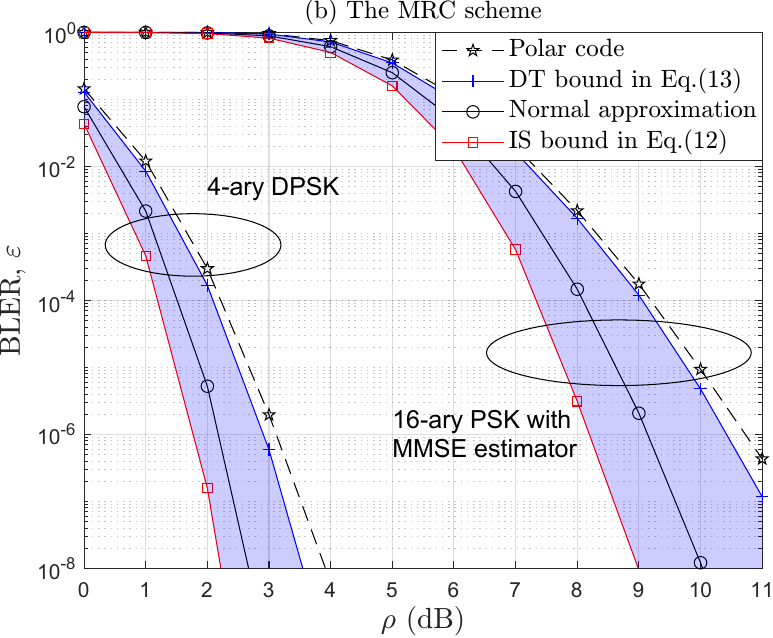}
}
\caption{BLER comparison of differential modulation and PSK with MMSE channel estimators under normalized Doppler value $f_dT_s =0.05$ and coding rate $R_c=0.5$.}
\end{figure*}
\setlength{\textfloatsep}{26pt}

Fig.6 illustrates the BLER performance of differential modulation over fading channels found by Corollary 1 for different combining schemes and numbers ($K$) of connected BSs.
For Fig.6, the transmitted packet size is 32 bytes.
It can be seen that, as expected, $\varepsilon$ of both the SC and MRC schemes decreases with increasing $K$.
Moreover, compared with the SC scheme, the MRC scheme has better BLER performance.
We also note that the analytical minimum achievable BLER of differential modulation can well describe the diminishing return problem of $K$ in the simulated BLER for both the SC and the MRC schemes, which verifies the correctness of our analysis.


Fig.7 and Fig.8 compare the BLER of differential modulation with that of PSK/QAM with MMSE channel estimators under different normalized Doppler values.
For a fair comparison, we assume that the net data rates and power of the DPSK and the corresponding PSK/QAM with MMSE channel estimators are the same.
Note that since the pilot symbols are used, the PSK/QAM transmission with MMSE channel estimators suffers from a capacity and energy loss compared to differential modulation.
Therefore, for obtaining a fair comparison, we need to carefully set all the relevant parameters of the modulation schemes.

\begin{figure*}[!ht]
\setlength{\abovecaptionskip}{5pt}
\setlength{\belowcaptionskip}{-5pt}
\centering
\begin{minipage}{1\linewidth}	
\centering
\subfloat{
\includegraphics[width=0.45\textwidth]{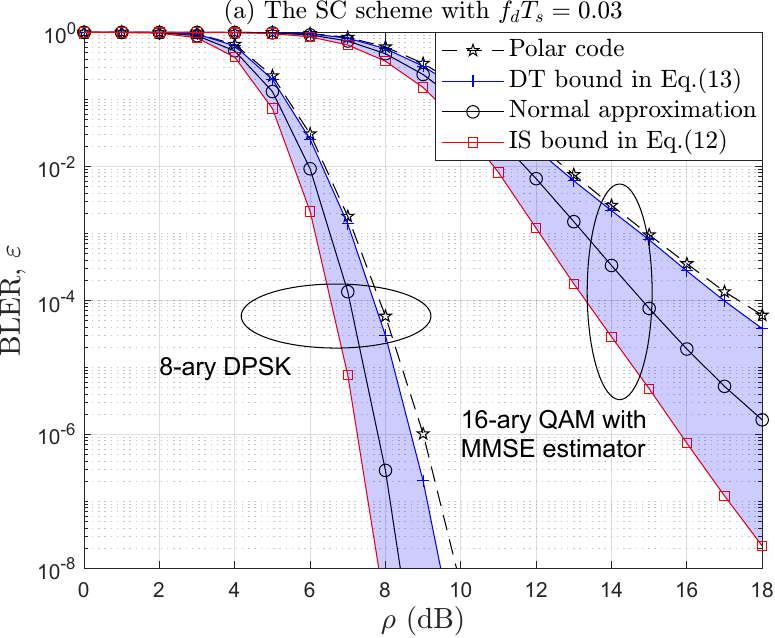}
}
\subfloat{
\includegraphics[width=0.45\textwidth]{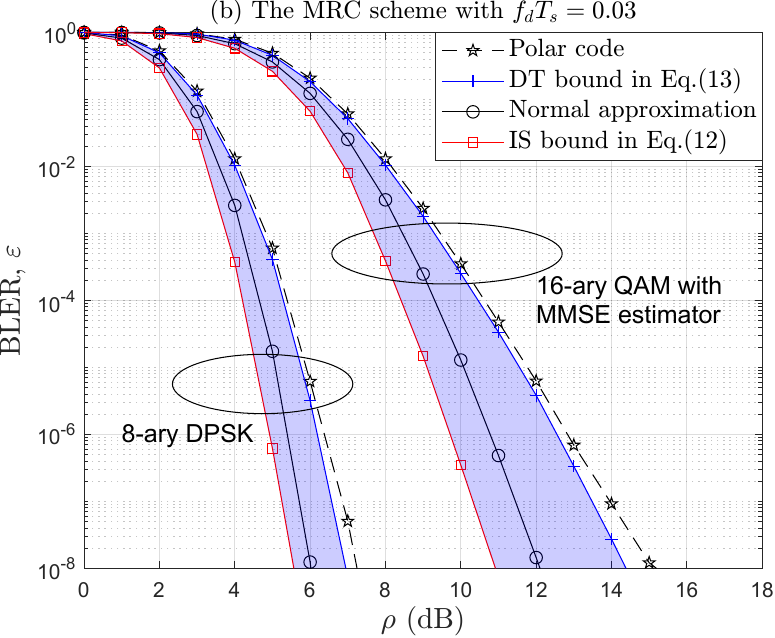}
}
\end{minipage}
\vskip 0cm 
\begin{minipage}{1\linewidth }
\centering
\subfloat{
\includegraphics[width=0.45\textwidth]{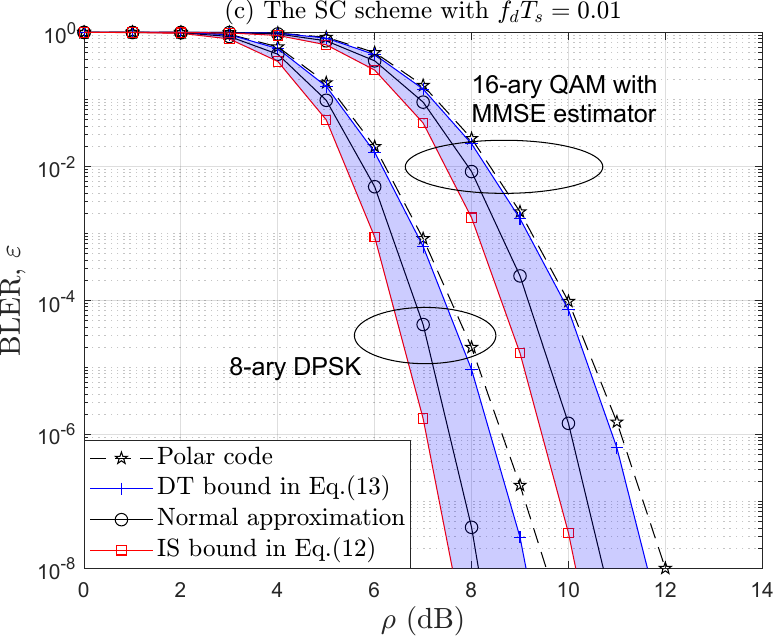}
}
\subfloat{
\includegraphics[width=0.45\textwidth]{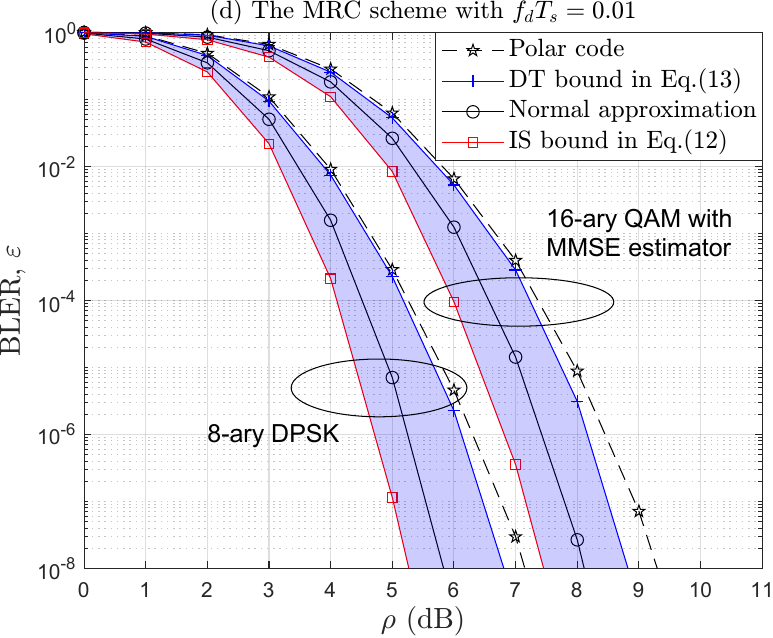}
}
\end{minipage}
\vspace{0pt}	
\caption{BLER comparison of differential modulation and QAM modulation with MMSE channel estimators under different normalized Doppler values.}
\end{figure*}

In Fig.7, the BLER performance of differentially modulated and PSK transmissions over fading channels determined by Corollary 1 and 2 is illustrated under normalized Doppler value $f_dT_s =0.05$.
Specifically, the number of the associated BSs is $K=3$ and the order of differential modulation is fixed at $M=4$.
In addition, to make the net data rate and energy consumption of the pilot-assisted scheme consistent with differential modulation,
the MMSE channel estimation system adopts PSK modulation with order $M=16$ and the pilot interval is $\upsilon =1$ (i.e. the most favourable setup for the coherent scheme!).
For a packet size of 32 bytes, the transmitted symbol blocklength of both the pilot-assisted transmission and differential transmission is $N=128$.
As a result, the power and net data rate of the pilot-assisted scheme are the same as that of differential modulation.
From the figure, the normal approximation, IS bound, DT bound and polar coding-based results show that differential modulation outperforms the pilot-assisted scheme with MMSE channel estimators in a high Doppler environment for both the SC and MRC schemes.
Thus, differential modulation does offer an advantage over the pilot-assisted scheme (even under the most favourable pilot setup) for SPT, especially in a high Doppler environment.

Fig.8 shows that the BLER performance of differentially modulated and QAM transmissions over fading channels determined by Corollary 1 and 3 under normalized Doppler values $f_dT_s =0.03$, $0.01$.
In particular, the number of the associated BSs is $K=3$, the order of differential modulation is $M=8$ and the packet size is set at $30$ bytes. For the simulated polar coding schemes, a (240, 120) polar code with the coding rate $R_c=0.5$ is assumed.
In addition, the MMSE channel estimation based QAM transmission adopts order $M=16$ and the pilot interval is $\upsilon =3$.
Therefore, the transmitted symbol blocklength of both the pilot-assisted transmission and the differential transmission is $N=80$. As expected, for a moderate to high Doppler environment, differential modulation shows a significant advantage over the pilot-assisted coherent scheme in terms of the normal approximation, IS bound, DT bound and polar coding-based BLER results.
Our results for higher Doppler values give even more advantage to differential modulation over the corresponding QAM transmission with the MMSE estimator.
In addition, it can be seen that $\varepsilon$ increases with increasing normalized Doppler value $f_dT_s$ for both differential modulation and the QAM modulation with MMSE channel estimation (as expected, but the differential modulation results are always much ahead).

\begin{figure*}[!ht]
\setlength{\abovecaptionskip}{5pt}
\setlength{\belowcaptionskip}{-5pt}
\centering
\subfloat{
\includegraphics[width=0.3\textwidth]{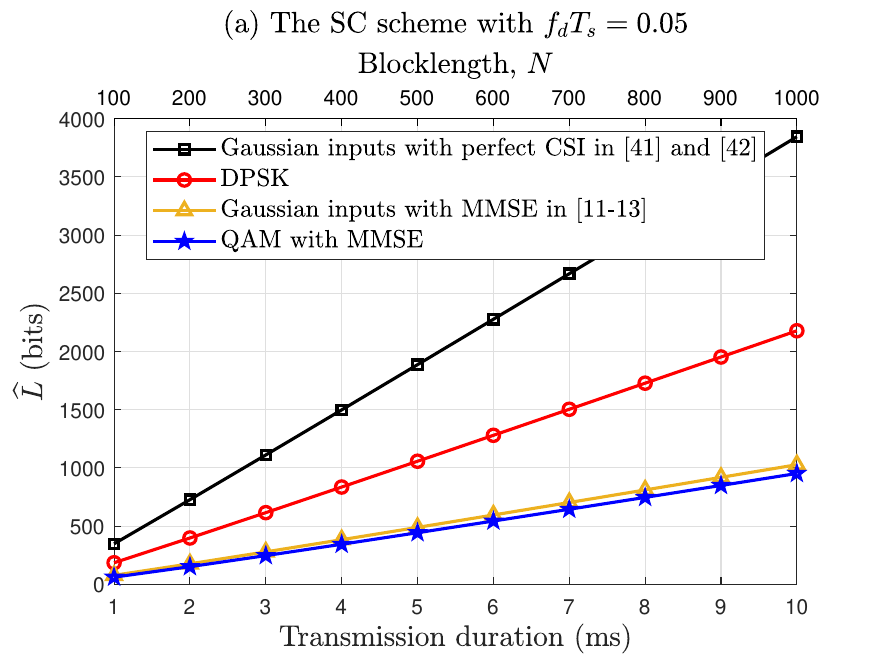}
}
\subfloat{
\includegraphics[width=0.3\textwidth]{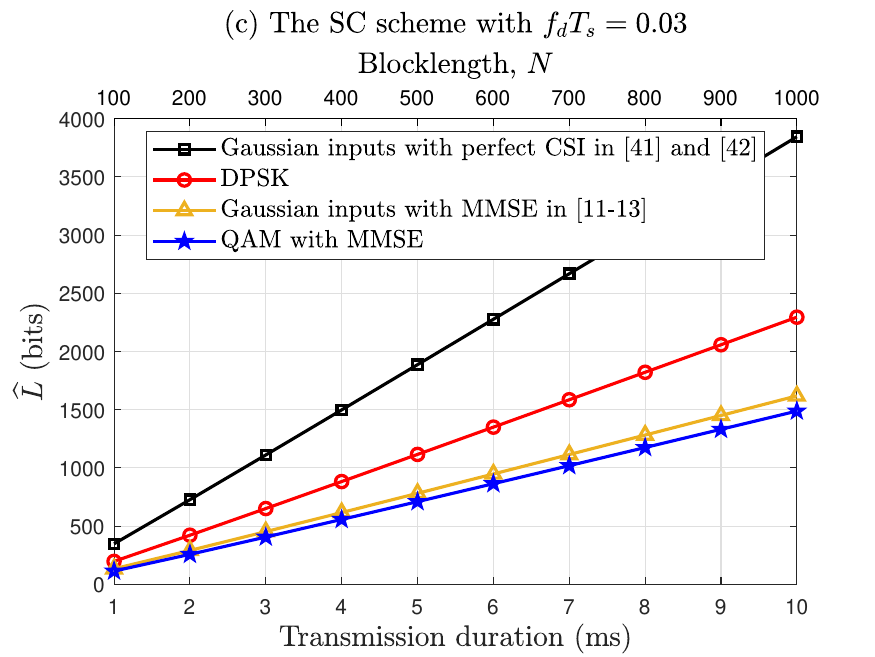}
}
\subfloat{
\includegraphics[width=0.3\textwidth]{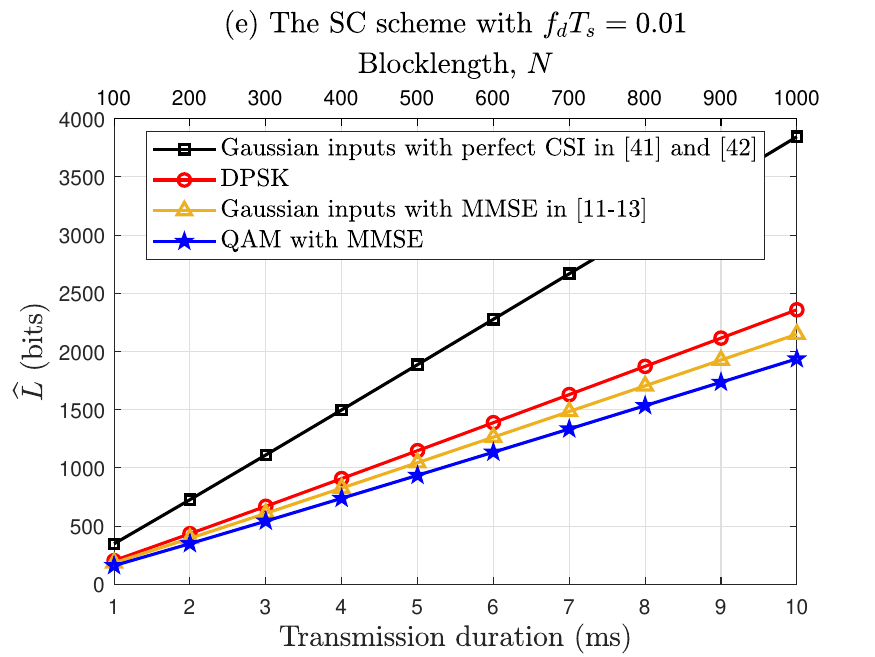}
}\\
\subfloat{
\includegraphics[width=0.3\textwidth]{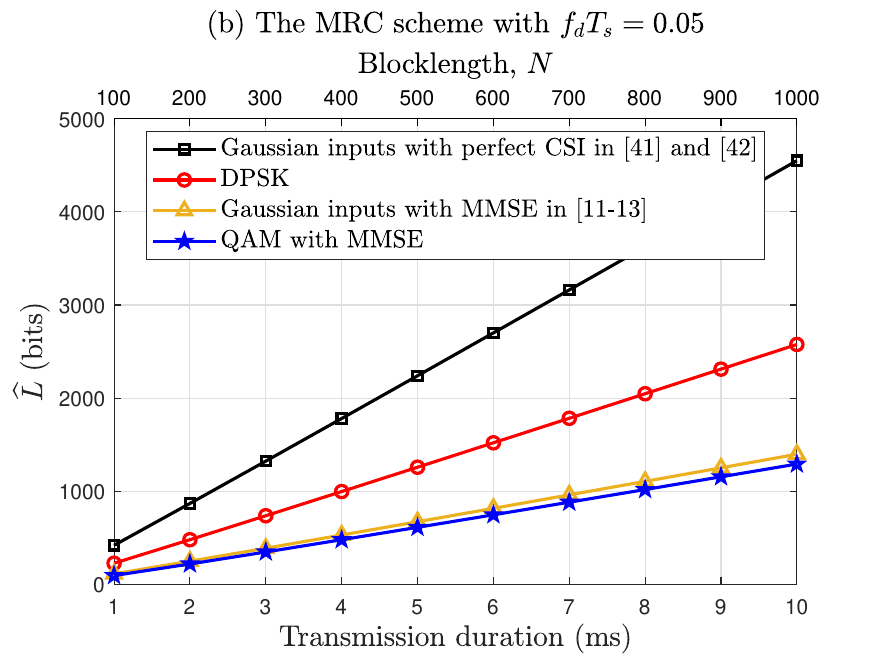}
}
\subfloat{
\includegraphics[width=0.3\textwidth]{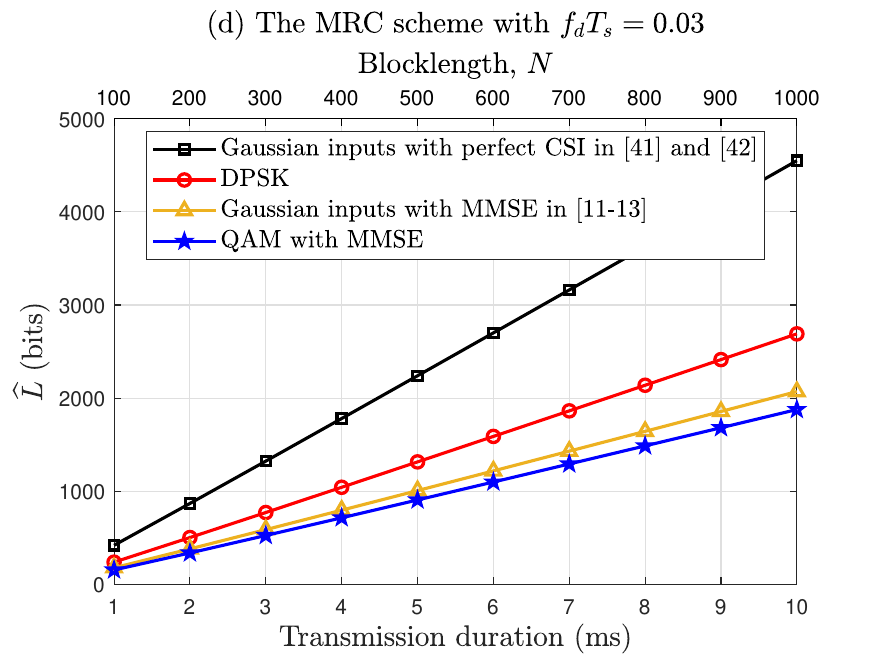}
}
\subfloat{
\includegraphics[width=0.3\textwidth]{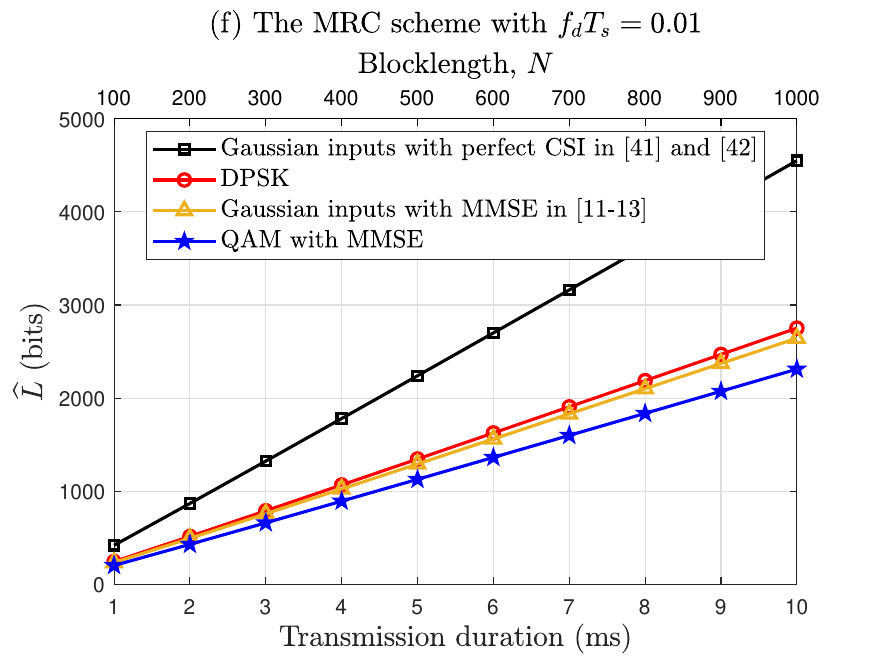}
}
\caption{Largest achievable information bit number comparison of Gaussian inputs with perfect CSI, DPSK, Gaussian inputs with MMSE and QAM modulation with MMSE under different normalized Doppler values.}
\end{figure*}
\setlength{\textfloatsep}{26pt}

Fig.9 illustrates a comparison of the largest information bit number $\widehat L$ for Gaussian inputs with perfect CSI from \cite{Polyanskiy2011Scalar} and \cite{Lancho2018Normal}, differential modulation in Eq.(\ref{rate_fbl_diff_fading}), Gaussian inputs with MMSE channel estimators from \cite{Johan2021URLLC,Lancho2023Cell,Kislal2023Efficient}, and QAM with MMSE in Eq.(\ref{rate_fbl_qam_mmse}) under different transmission durations, block lengths, and normalized Doppler values.
The system bandwidth in Fig.9 is set to be 100 kHz, resulting in a symbol duration of $T_s = 0.01$ ms (assuming zero roll-off).
The corresponding block lengths for different transmission durations are displayed on the top axis of the figures.
Also, both differential modulation and QAM are assumed to be of the same order, $M=16$, as they have identical net data rates and power for each transmission duration (block length).
In addition, the following parameters are used: the reliability requirement $\varepsilon= 10^{-5}$, SNR $\rho=10$dB, $K=3$.
It can be seen that Gaussian inputs with perfect CSI achieve the highest information payload in SPT for various transmission durations, block lengths, and normalized Doppler values.
Furthermore, it is evident that, for the same latency and reliability requirements, differential modulation always outperforms the corresponding pilot-assisted scheme in terms of information payload in the SPT.
This advantage stems from two main factors: (1) differential modulation needs no channel estimation and thus far less sensitive to channel variation or drifting, and (2) it does not incur pilot overhead,  resulting in a relatively higher net data rate (or shorter transmission duration or latency) under the same modulation level and transmit power (or equivalently a lower modulation level or lower BLER under the same net data rate and transmit power, as shown earlier).
Additionally, we observe that $\widehat L$ of the pilot-assisted scheme experiences a significant decrease as $f_dT_s$ increases for both Gaussian inputs and QAM inputs.
However, the influence of $f_dT_s$ on $\widehat L$ is much less pronounced for differential modulation.
This discrepancy can be attributed to the fact that larger $f_dT_s$ results in faster channel drifting (thus leading to poorer channel accuracy for payload symbols) in the pilot-assisted scheme, while the differential transmission is far less sensitive to channel drifting.
\section{Conclusion}

The benefits of jointly differential modulation and MC schemes in SPT has been investigated in this paper.
Thanks to differential modulation, no pilots and CSI are needed at the receiver for detection of the transmitted symbols, and thus the power and bandwidth overhead incurred in channel estimation schemes is eliminated (leading to lower transmission and processing latencies).
By employing the non-asymptotic information-theoretic bounds, we first derived the maximum achievable rate and minimum achievable BLER performance of point-to-point i.i.d ergodic fading channels with i.i.d PSK inputs.
Then, the maximum achievable rate and minimum achievable BLER performance results are extended to differential modulation and MC schemes.
Numerical and simulated results demonstrate that, differential modulation outperforms the pilot-assisted coherent counterpart, especially in a moderate and high Doppler environment. Although only a first attempt, it reveals the huge potential of adopting differential modulation for SPT in URLLC.


In addition, the derived SPT performance results with differential modulation are useful in two ways.
Firstly, the proposed results extend the normal approximation provided in \cite{Polyanskiy2010Channel}, \cite{Polyanskiy2011Scalar} and \cite{Lancho2018Normal}, whose channel inputs are assumed to be continuous.
Thus, the obtained results in this paper are more suitable to analyze the SPT performance with practical modulation schemes and discrete inputs.
Secondly, the results provide the foundation for analytical studies that analyze the behavior of the maximum transmission rate or the minimum BLER as a function of system parameters such as SNR, blocklength and modulation order.

\ifCLASSOPTIONcaptionsoff
  \newpage
\fi
\section*{Appendix A: proof of Theorem 3}
The proof mainly relies on the Berry-Esseen Theorem \cite[Ch. XVI.5]{An1971William}.
Specifically, the Berry-Esseen Theorem states that the probability that the sum of $N$ independent random variables $\left\{ {{X_n}} \right\}_{n = 1}^N$ with expectation ${I_n}$, variance $V_n$ and absolute third moment $T_n$ taking value over some interval can be well described by the first and second order moments of the random variables and for any $ \lambda $ :
\begin{equation}
\begin{aligned}
\left| {P\left[ {\frac{{\sum\nolimits_{n = 1}^N {\left( {{X_n} - {I_n}} \right)} }}{{\sqrt {NV} }} \geqslant \lambda } \right] - Q\left( \lambda  \right)} \right| \leqslant \frac{{6T}}{{{V^{3/2}}\sqrt N }}
\label{proof_Th2_BE_Theo}
\end{aligned}
\end{equation}
where $V = \sum\nolimits_{n = 1}^N {\frac{{{V_n}}}{N}} $ and $T = \sum\nolimits_{n = 1}^N {\frac{{{T_n}}}{N}} $.

It is well known that the conditional transition probability ${{\mathbb{P}}_{Y^{N}\left| X^{N} \right.}}\left( {y^{N}\left| x^{N} \right.} \right)= \prod\nolimits_{n = 1}^N {{\mathbb{P}}_{Y\left| X \right.}}\left( {y_n\left| x_n \right.} \right)$ is valid for i.i.d ergodic fading channels.
Then, the expectation and variance of the information density with blocklength $N$ for i.i.d ergodic fading channels can be expressed as
$\mathbb{E} \left[ i\left( {X^{N};Y^{N}} \right) \right] = \sum\nolimits_{n = 1}^N {\mathbb{E} \left[ i\left( {X^{n};Y^{n}} \right) \right]} = N I_{\text{coh}}\left( \rho \right)$ and
$\mathbb{V}\text{ar} \left[ i\left( {X^{N};Y^{N}} \right) \right] = \sum\nolimits_{n = 1}^N {\mathbb{V}\text{ar} \left[ i\left( {X^{n};Y^{n}} \right) \right]} = N V_{\text{coh}}\left( \rho \right)$.
By applying the Berry-Esseen Theorem to $i\left( {{X^N};{Y^N}} \right)$, we further have
\begin{equation}
\begin{aligned}
&\left| {P\left[ {\frac{   i\left( {{X^N};{Y^N}} \right) - N I_{\text{coh}}\left( \rho \right)  }{{\sqrt {{NV_{\text{coh}}\left( \rho \right)}} }} \geqslant \lambda } \right] - Q\left( \lambda  \right)} \right|\\
&\mathop  = \limits^{(a)}\left| {P\left[ {i\left( {{X^N};{Y^N}} \right) \leqslant N I_{\text{coh}}\left( \rho \right)-\mu {{\sqrt {NV_{\text{coh}}\left( \rho \right)} }} } \right] - Q\left( \mu  \right)} \right| \\
&\leqslant \frac{{6T}}{{{(V_{\text{coh}}\left( \rho \right))^{3/2}}\sqrt N }},
\label{proof_Th2_BE_Theo_deri1}
\end{aligned}
\end{equation}
where $(a)$ is obtained by letting $\lambda =-\mu$ and using $Q(x)=1-Q(-x)$.

Based on the result in (\ref{proof_Th2_BE_Theo_deri1}), we can obtain the maximum achievable rate and the minimum achievable BLER under i.i.d erdodic fading channels by respectively employing the IS and DT bounds.

\textbf{A.1 IS bound}

By considering $\widehat L$ information bits in the information spectrum bound of Theorem 1, we have
\begin{equation}
\begin{aligned}
\varepsilon  &\geqslant \mathop {\sup }\limits_{\beta  > 0} \left\{ {\mathop {\inf }\limits_{{{\mathbb{P}}_X}} {\mathbb{P}}\left[ {i\left( {X^{N};Y^{N}} \right) \leqslant {{\log }_2}\beta } \right] - \frac{\beta }{{{2^{\widehat L}}}}} \right\}\\
&\mathop  \geqslant \limits^{\left( a \right)}  {\mathbb{P}}\left[ {{i\left( {X^{N};Y^{N}} \right) } \leqslant {{\log }_2}\left( {\frac{{{2^{\widehat L}}}}{{\sqrt N }}} \right)} \right] - \frac{1}{{\sqrt N }}\\
&\mathop  \geqslant \limits^{\left( b \right)} Q\left( {\frac{{N I_{\text{coh}}\left( \rho \right) - {{\log }_2}\left( {{{{2^{\widehat L}}} \mathord{\left/
 {\vphantom {{{2^{\widehat L}}} {\sqrt N }}} \right.
 \kern-\nulldelimiterspace} {\sqrt N }}} \right)}}{{\sqrt {NV_{\text{coh}}\left( \rho \right)} }}} \right) - \frac{B}{{\sqrt N }},
\label{proof_Th3_IS}
\end{aligned}
\end{equation}
where $B=\frac{{6T}}{{{(V_{\text{coh}}\left( \rho \right))^{3/2}} }}-1$,
$(a)$ follows from the change of variables $\beta  = \frac{{{2^{\widehat L}}}}{{\sqrt N }}$ and the fact from \cite{Ungerboeck1982Channel} and \cite{A2014Jazi} that the discrete uniform distribution, namely i.i.d. PSK/QAM input distribution, is capacity-achieving input distributions for channels with $M$-ary PSK/QAM inputs,
while $(b)$ is obtained by using the result of (\ref{proof_Th2_BE_Theo_deri1}) and letting $\mu={\frac{{N I_{\text{coh}}\left( \rho \right) - {{\log }_2}\left( {{{{2^{\widehat L}}} \mathord{\left/
 {\vphantom {{{2^{\widehat L}}} {\sqrt N }}} \right.
 \kern-\nulldelimiterspace} {\sqrt N }}} \right)}}{{\sqrt {NV_{\text{coh}}\left( \rho \right)} }}}$.
Based on the result of (\ref{proof_Th3_IS}), the maximum achievable rate can be further derived as
\begin{equation}
\begin{aligned}
&{\widehat R}\left( {N,\varepsilon } \right)= \frac{\widehat L}{N}\\
&\leqslant I_{\text{coh}}\left( \rho \right) - \sqrt {\frac{V_{\text{coh}}\left( \rho \right)}{N}} {Q^{ - 1}}\left( {\varepsilon  + \frac{B}{{\sqrt N }}} \right) + \frac{1}{2}{\log _2}N\\
&\mathop  = \limits^{\left( a \right)}  I_{\text{coh}}\left( \rho \right) - \sqrt {\frac{V_{\text{coh}}\left( \rho \right)}{N}} {Q^{ - 1}}\left( {\varepsilon} \right) + \frac{{\log _2}N}{2N}+\mathcal{O}\left( \frac{1}{n} \right),
\label{proof_Th3_rate_upper}
\end{aligned}
\end{equation}
where $(a)$ follows from the Taylor expansion for the ${Q^{ - 1}}\left( {\cdot} \right)$ function.

\textbf{A.2 DT bound}

By considering $\widehat L$ information bits in the DT bound of Theorem 2, we have
\begin{equation}
\begin{aligned}
&\varepsilon  \leqslant \mathbb{E}\left[ {\exp \left\{ { - {{\left[ {i\left( {{X^N};{Y^N}} \right) - {{\log }_2}\left( {\frac{{{2^{\widehat L}} - 1}}{2}} \right)} \right]}^ + }} \right\}} \right]\\
& \mathop  = \limits^{(a)}\underbrace {\mathbb{P}\left[ i\left( {{X^N};{Y^N}} \right)  \leqslant {{\log }_2}\left( {\frac{{{2^{\widehat L}} - 1}}{2}} \right) \right]}_{\text{Term~1}}+\\
&\underbrace {{\frac{{{2^{\widehat L}} - 1}}{2}} \mathbb{E}\left[ {\exp \left\{ { - {{ {i\left( {{X^N};{Y^N}} \right) } }}} \right\}} \mathbbm{1}_{\left\{ {i\left( {{X^N};{Y^N}} \right) - {{\log }_2}\left( {\frac{{{2^{\widehat L}} - 1}}{2}} \right)} \right\}} \right]}_{\text{Term~2}},
\label{proof_Th2_DT}
\end{aligned}
\end{equation}
where $ \mathbbm{1}_{\left\{ x \right\}}$ is the indicator function and $(a)$ is obtained by using the definition of ${\left[ A \right]^ + }$ in the Theorem 2 and dividing the expectation operation into two case, $i\left( {{X^N};{Y^N}} \right)  \leqslant {{\log }_2}\left( {\frac{{{2^{\widehat L}} - 1}}{2}} \right) $ and $i\left( {{X^N};{Y^N}} \right)  > {{\log }_2}\left( {\frac{{{2^{\widehat L}} - 1}}{2}} \right) $.
Next, we separately analyse these two terms in (\ref{proof_Th2_DT}).


For the first term of (\ref{proof_Th2_DT}),
we set
\begin{equation}
\begin{aligned}
{{\log }_2}\left( {\left({{ {2^{\widehat L}} - 1}}\right)/{2}} \right) =N I_{\text{coh}}\left( \rho \right)-\mu {{\sqrt {NV_{\text{coh}}\left( \rho \right)} }}.
\label{proof_Th2_term1_deri}
\end{aligned}
\end{equation}
Then, by substituting (\ref{proof_Th2_term1_deri}) into (\ref{proof_Th2_BE_Theo_deri1}), the first term of (\ref{proof_Th2_DT}) can be written as
\vspace{-0.5em}
\begin{equation}
\begin{aligned}
&{\mathbb{P}\left[ i\left( {{X^N};{Y^N}} \right)  \leqslant {{\log }_2}\left( {\left({{ {2^{\widehat L}} - 1}}\right)/{2}} \right) \right]}\leqslant Q\left( \mu  \right) +\\
&{{6T}}{{{(V_{\text{coh}}/\left( \rho \right))^{3/2}}\sqrt N }}.
\label{proof_Th2_term1}
\end{aligned}
\end{equation}
In addition, by using the derivation techniques in Appendix G of \cite{Polyanskiy2010Channel}, the second term of (\ref{proof_Th2_DT}) can be derived as
\begin{equation}
\begin{aligned}
&{{\frac{{{2^{\widehat L}} - 1}}{2}} \mathbb{E}\left[ {\exp \left\{ { - {{ {i\left( {{X^N};{Y^N}} \right) } }}} \right\}} \mathbbm{1}_{\left\{ {i\left( {{X^N};{Y^N}} \right) - {{\log }_2}\left( {\frac{{{2^{\widehat L}} - 1}}{2}} \right)} \right\}} \right]}\\
&\leqslant 2\left( {\frac{{{{\log }_2}2}}{{\sqrt {2\pi } }} + \frac{{12T}}{{V_{\text{coh}}\left( \rho  \right)}}} \right)\frac{1}{{\sqrt {NV_{\text{coh}}\left( \rho  \right)} }}.
\label{proof_Th2_term2}
\end{aligned}
\end{equation}

Note that the inequality in (\ref{proof_Th2_DT}) can be established by summing (\ref{proof_Th2_term1}) and (\ref{proof_Th2_term2}), and selecting a suitable $\mu$ as follows
\begin{equation}
\begin{aligned}
&\mu= Q^{-1}\left( \varepsilon - \left( {\frac{{{2{\log }_2}2}}{{\sqrt {2\pi } }} + \frac{{30T}}{{V_{\text{coh}}\left( \rho  \right)}}} \right)\frac{1}{{\sqrt {NV_{\text{coh}}\left( \rho  \right)} }} \right)
\label{proof_Th2_DT_summ}
\end{aligned}
\end{equation}

Furthermore, let $\varepsilon_{Q}=\varepsilon - \left( {\frac{{{2{\log }_2}2}}{{\sqrt {2\pi } }} + \frac{{30T}}{{V_{\text{coh}}\left( \rho  \right)}}} \right)\frac{1}{{\sqrt {NV_{\text{coh}}\left( \rho  \right)} }}$. By utilizing the Taylor expansion to $\varepsilon$, (\ref{proof_Th2_DT_summ}) can be approximated as
\begin{equation}
\begin{aligned}
Q^{-1}\left( \varepsilon_{Q} \right)= Q^{-1}\left( \varepsilon \right) + \left( {\frac{d Q^{-1}\left( \varepsilon \right)}{{d\varepsilon }} + \mathcal{O}\left( 1 \right) } \right)\left( {\varepsilon_{Q} -\varepsilon } \right)
\label{proof_Th2_Tay}
\end{aligned}
\end{equation}

Finally, by substituting (\ref{proof_Th2_Tay}) into (\ref{proof_Th2_term1_deri}), we can obtain
\begin{equation}
\begin{aligned}
{\widehat L} \geqslant {\log _2}\frac{{{2^{\widehat L}} - 1}}{2}
= N I_{\text{coh}}\left( \rho \right)- {{\sqrt {NV_{\text{coh}}\left( \rho \right)} }}Q^{-1}\left( \varepsilon \right)+\mathcal{O}\left( 1 \right).
\label{proof_Th2_rate2}
\end{aligned}
\end{equation}

Note that the upper bound in (\ref{proof_Th3_rate_upper}) and the lower bound in (\ref{proof_Th2_rate2}) are approximately equivalent and thus are tight. Then, the maximum achievable rate can be approximated as
${\widehat R}\left( {N,\varepsilon } \right) \approx   I_{\text{coh}}\left( \rho \right) - \sqrt {\frac{V_{\text{coh}}\left( \rho \right)}{N}} {Q^{ - 1}}\left( {\varepsilon} \right)+ \frac{{\log _2}N}{2N}$.
$\hfill\blacksquare$

\end{document}